\documentclass[iop,numberedappendix]{emulateapj}

\usepackage{amsmath}
\usepackage{natbib}

\newcommand{\kms}{km s$^{-1}$}
\newcommand{\msun}{M$_{\odot}$}

\shorttitle{ALFALFA HI Mass Function}
\shortauthors{Martin et al.}

\begin{document}

\title{The Arecibo Legacy Fast ALFA Survey: X. The HI Mass Function and $\Omega_{HI}$ from the 40\% ALFALFA Survey}

\author {Ann M. Martin\altaffilmark{1}, Emmanouil Papastergis\altaffilmark{1},
Riccardo Giovanelli\altaffilmark{1,2}, Martha P. Haynes\altaffilmark{1,2}, Christopher M. Springob\altaffilmark{3}, Sabrina Stierwalt\altaffilmark{4}}
\altaffiltext{1}{Center for Radiophysics and Space Research, Space Sciences Building,
Cornell University, Ithaca, NY 14853. {\textit{e-mail:}} amartin@astro.cornell.edu,
papastergis@astro.cornell.edu,riccardo@astro.cornell.edu, haynes@astro.cornell.edu}
\altaffiltext{2}{National Astronomy and Ionosphere Center, Cornell University,
Ithaca, NY 14853. The National Astronomy and Ionosphere Center is operated
by Cornell University under a cooperative agreement with the National Science
Foundation.}
\altaffiltext{3}{Australian Astronomical Observatory, P.O. Box 296, Epping, NSW 1710, Australia. {\textit{e-mail:}} springob@aao.gov.au}
\altaffiltext{4}{Spitzer Science Center, California Institute of Technology, 1200 E. California Blvd. 91125. {\textit{e-mail:}} sabrina@ipac.caltech.edu}

\begin{abstract}
The Arecibo Legacy Fast ALFA (ALFALFA) survey has completed source extraction for 40$\%$ of its total sky area, resulting in the largest sample of HI-selected galaxies to date. We measure the HI mass function from a sample of 10,119 galaxies with $ 6.2 < \log (M_{HI}/M_{\odot}) < 11.0$ and with well-described mass errors that accurately reflect our knowledge of low-mass systems.  We characterize the survey sensitivity and its dependence on profile velocity width, the effect of large-scale structure, and the impact of radio frequency interference in order to calculate the HIMF with both the 1/V$_{max}$ and 2DSWML methods. We also assess a flux-limited sample to test the robustness of the methods applied to the full sample. These measurements are in excellent agreement with one another; the derived Schechter function parameters are $\phi_{*}$ (h$_{70}^3$ Mpc$^{-3}$dex$^{-1}$) = 4.8 $\pm$ .3 $\times$ 10$^{-3}$, $\log $(M$_{*}/$M$_{\odot})$ + 2 $\log$ h$_{70}$  = 9.96 $\pm$ 0.02 and $\alpha$ = -1.33 $\pm$ 0.02. We find $\Omega_{HI} =$ 4.3 $\pm$ 0.3 $\times 10^{-4}$ h$^{-1}_{70}$, 16$\%$ larger than the 2005 HIPASS result, and our Schechter function fit extrapolated to $\log (M_{HI}/M_{\odot})$ = 11.0 predicts an order of magnitude more galaxies than HIPASS. The larger values of $\Omega_{HI}$ and of M$_{*}$ imply an upward adjustment for estimates of the detection rate of future large-scale HI line surveys with, e.g., the Square Kilometer Array. A comparison with simulated galaxies from the Millennium Run and a treatment of photoheating as a method of baryon removal from HI-selected halos indicates that the disagreement between dark matter mass functions and baryonic mass functions may soon be resolved.
\end{abstract}

\section{Introduction \label{intro}}

The disagreement between predictions of the number of low-mass dark matter halos and the observations of low-luminosity dwarf galaxies, commonly characterized as the `missing satellite problem,' is reflected in the faint-end slopes of galaxy luminosity functions and neutral hydrogen (HI) mass functions. Current dark matter simulations and models \citep{2009MNRAS.398.1150B,2001MNRAS.321..372J} imply that the faint-end slope of the underlying mass function is $\alpha \sim -1.8$, in agreement with the Press-Schechter analysis of cosmic structure formation \citep{1974ApJ...187..425P}, but observational evidence is consistent with a significantly shallower slope.

There is hope of resolving this discrepancy by investigating physical effects on the observed baryons that would not influence the underlying dark matter distribution. For example, photoheating by the UV background can deplete baryons from low mass halos, reducing the number of luminous galaxies observable today. There appears to be a characteristic halo mass, below which severe baryon depletion could eliminate the abundance of dwarf galaxies \citep{2008IAUS..244..279H}; \citet{2010arXiv1001.4721H} find this halo mass to be $\approx 6 \times 10^9h^{-1}$ \msun, and that it is robust against assumed UV background flux density and simulation resolution effects. Other processes related to star formation, such as supernova feedback \citep{2000MNRAS.317..697E} can remove gas from galaxies, preferentially removing baryons from those early galaxies residing in weak potential wells. Understanding these baryonic processes has the potential to resolve the missing satellite problem \citep{2007ApJ...670..313S}, but it remains difficult to fully simulate baryons in forming and evolving galaxies \citep{2007MNRAS.374.1479G,2008ASL.....1....7M, 2009ApJ...695..292C,2009ApJ...697...55G}, and it is therefore important to develop other observational constraints. 

Since low-mass dark matter halos are the most likely to suffer from baryon depletion, these effects may cause the shallow faint-end slopes observed in luminosity, circular velocity \citep{2009arXiv0912.1754Z}, and HI mass functions (HIMFs). Detailed study of these influences in the lowest-mass galaxies are only possible very nearby, and the dwarf galaxies in the Local Group have been shown to have great diversity in their star formation histories and metallicities \citep{2009ARA&A..47..371T,2004ApJ...610L..89G}, with some galaxies losing gas and ceasing star formation early while others have undergone this process only recently. Recently, \citet{2009MNRAS.392L..45R} has suggested that these halos may be able to re-accrete cold gas at late times, and proposes that the gas-bearing ultrafaint dwarf Leo T \citep{2007ApJ...656L..13I,2008MNRAS.384..535R} may be an example of this process. Such galaxies may then be observable in HI line surveys like the ALFALFA survey \citep{2010ApJ...708L..22G}. 

Blind HI surveys are ideal for probing these questions surrounding the lowest-baryon systems. HI line surveys are unbiased by properties like optical surface brightness, and ALFALFA in particular is designed to detect systems with lower HI masses than the blind surveys of the previous generation, down to $\sim 3 \times 10^{7}$ \msun \ at the distance of the Virgo cluster with SNR $\sim$ 6.5 \citep{2007AJ....133.2569G}. Since neutral gas fractions become large for dwarf galaxies, dominating the stellar mass, HI surveys are efficient at finding the extremely low-baryon-mass systems locally \citep{2001AJ....121.2420S,2006ApJ...653..240G}, and the HIMF is a better measure of baryon content at the lowest masses. Furthermore, environment is well known to have an impact on gas reservoirs, with galaxies in clusters tending to be HI deficient compared to those in the field \citep{1984ARA&A..22..445H}. The results of this bias as seen in the ALFALFA survey catalogs and in HI mass functions of various environments may provide insights to the relationship between HI gas densities, tidal and ram pressure stripping, and star formation.

Surveys like ALFALFA which probe a cosmologically fair sample also provide a wealth of information on the rare galaxies at the highest masses. High-mass gas-rich galaxies constrain the cosmic density of neutral gas in the local universe, $\Omega_{HI}$. HI contributes only about 1$\%$ of the baryon budget at z=0 \citep{2009and..book..419P,2004ApJ...616..643F,1998ApJ...503..518F}. The HI mass function is necessary to estimate this with great precision in order to trace the evolution of the neutral gas fraction, measured through damped Ly$\alpha$ systems at higher redshifts.

HI surveys also have the advantage of combining a galaxy detection, a redshift, and a mass estimate in a single observation without followup.  This is particularly important given that about 70$\%$ of galaxies in the blind ALFALFA catalog are new HI detections and many are altogether new redshifts, indicating that the conventional wisdom guiding targeted HI surveys toward galaxies expected to contain large reservoirs was severely limited. Finally, as simulations and semianalytic models of warm and cold gas in evolving galaxies improve, the HIMF can be used as a test of these results, as done in \citet{2009ApJ...698.1467O} through a comparison of modeled cold hydrogen gas in Millennium Run galaxies to the \citet{2005MNRAS.359L..30Z} mass function (see \S \ref{simulations}).

The first generation of blind HI surveys resulting in a measurement of the local HIMF contained few galaxies: \citet{2000AJ....119.2686H} detected 110 galaxies in the Southern Zone of Avoidance, and the Arecibo Dual Beam Survey (ADBS) HIMF was based on a sample of 265 galaxies \citep{2002ApJ...567..247R}. Both found a faint-end slope $\alpha \sim -1.5$, significantly steeper than what is found in other larger blind HI surveys. The published HIPASS HIMFs were based on more galaxies than previous blind surveys; the function extracted from the 1000 brightest detections \citep{2003AJ....125.2842Z} had a faint-end slope -1.3 and the later paper, with a fuller catalog of 4315 sources \citep{2005MNRAS.359L..30Z}, found -1.37. At the low-mass end of the HIMF, there is clearly severe disagreement, and previous data did not include enough low-mass objects to robustly constrain masses $< 10^{8}$ \msun. \citet{2005ApJ...621..215S} investigated a complete sample of 2771 optically-selected galaxies and found a shallow slope, $\alpha \sim -1.24$. Improving the number of sources by, for example, increasing the area of a shallow survey is not enough, on its own, to resolve the issue; rather, increasing the volume over which low-mass sources are detectable has the largest impact. Distance uncertainties are largest nearby, so a shallower survey will tend to base its low-mass slope on more uncertain objects \citep{2004ApJ...607L.115M}. 

The ALFALFA survey catalogs, including those previously published \citep{2007AJ....133.2569G, 2008AJ....135..588S, 2008AJ....136..713K,2009AJ....138..338S, 2009ApJS..183..214M} and those about to be published (Haynes et al. 2010, in prep), now represent $\sim40\%$ of the final survey area, and the HI mass function presented here considers a sample of $\sim 10000$ HI-selected galaxies. In the following section, we will discuss the ALFALFA dataset (\S2). In \S \ref{corrs} and \S \ref{swml}, respectively, we describe the 1/V$_{max}$ method of estimating the HIMF from corrected galaxy counts, and the two-dimensional Stepwise Maximum Likelihood (2DSWML) method. Details of these methods are discussed in Appendices A and B. After presenting the results of the global measurement of the HIMF along with $\Omega_{HI}$ in \S4 and 5, we will discuss the results as compared to the expectations of dark matter simulations and those including cold gas, addressing the divergence between HIMF slopes and that predicted by the Press-Schechter formalism (\S6). 

\section{ALFALFA Dataset \label{dataset}}

\subsection{The ALFALFA Survey \label{alfalfa}}
The ongoing ALFALFA survey takes advantage of the new multipixel ALFA receiver at the Arecibo Observatory. When complete, the survey will have measured $> 30,000$ galaxies in the 21 cm line out to $z \sim 0.06$ with a median redshift of $\sim$ 8000 \kms. The survey is more sensitive than HIPASS, with a 5$\sigma$ detection limit of 0.72 Jy \kms \ for a source with profile width 200 \kms \ in ALFALFA compared to a 5$\sigma$ sensitivity 5.6 Jy \kms \ for the same source in HIPASS \citep{ 2005AJ....130.2613G}. Narrow profile widths, down to $\sim$ 15 \kms, allow us to probe extremely small objects. ALFALFA detects objects with neutral hydrogen masses $M_{HI} \sim 3 \times 10^{7}$ \msun \ out to the distance of the Virgo cluster. In addition to greater sensitivity, ALFALFA probes gas-rich galaxies in the local universe with greater velocity resolution (11 \kms after Hanning smoothing \ vs. 18 \kms) and a deeper limiting redshift (18000 \kms \ vs. 12700 \kms) than HIPASS. Our significantly improved survey depth for low-mass objects allows the ALFALFA survey to better constrain the low-mass slope of the HI mass function.

ALFALFA survey data are acquired in a minimally-invasive drift scanning mode, in two passes ideally separated by several months, and individual 600 s drift scans are combined into three-dimensional data grids covering 2.4$^{\circ}$ in both R.A. and decl.; it therefore takes many nights of observations to complete a grid from which extragalactic sources can be extracted. 

Confidently detected sources are assigned one of three object codes, where Code 1 refers to a reliable extragalactic detection with a high S/N ($> 6.5$), Code 2 refers to extragalactic sources with marginal S/N ($4.5 < S/N < 6.5$) confirmed by an optical counterpart with known optical redshift matching the HI measurement, and Code 9 refers to High Velocity Clouds (HVCs). For this analysis, we consider only objects designated Code 1, since we are interested in extragalactic objects with well-known selection criteria. Code 1 objects have a reliable S/N, a good match between the two polarizations that are independently observed by ALFALFA, a clean spectral profile and, in almost every case, a confident match with an optical counterpart. The signal detection pipeline, discussed at length in \citet{2007AJ....133.2087S}, combines a matched-filtering technique for identifying source candidates with an interactive process for source confirmation and parameter measurement. This technique is estimated to result in a reliability of candidate detections $\sim 95\%$ for Code 1 objects, with a completeness better than $90 \%$ for the narrowest galaxies above the prescribed S/N threshold. The subsample of Code 1 objects provides a robust sample for the HIMF.

\subsection{Derived Parameters \label{derived}}

Published ALFALFA catalogs contain a set of measured parameters (including coordinates, heliocentric velocity, line profile velocity width W50 measured at the 50$\%$ level of two profile peaks, integrated flux density $S_{int}$, S/N, and noise figure $\sigma_{rms}$) in addition to a distance estimate and a derived HI mass in solar units, obtained from the expression $M_{HI} = 2.356 \times 10^{5} D^{2}_{Mpc} S_{int}.$ Our distance estimates are subject to errors due to each galaxy's unknown peculiar velocity, which translate into mass errors. The fractional distance error due to peculiar velocity decreases with increasing distance (the so-called `Eddington effect'), so the lowest-mass galaxies which are only found nearby are most prone to this error, our treatment of which is discussed in detail in \S \ref{disterrs}.

\subsection{Profile Width-Dependent Sensitivity \label{width}}

ALFALFA's ability to detect a signal depends not only on the integrated flux, but also on the profile width W$_{50}$ (\kms). Fig. \ref{sensitivity} displays the distribution of sources detected by ALFALFA. Rather than a single flux limit, the ALFALFA detection threshold is dependent on both S$_{int}$ and profile width W$_{50}$, and we find that this relationship changes above W$_{50} \sim$400 \kms. We fit the S$_{int, th}$ relationship empirically to the data, rather than using the assumed expression above. Due to differences in the two methods we employ to calculate the HIMF, we consider two different threshold cuts, described separately in \S \ref{corrs} and \ref{swml}.

\subsection{The 40\% ALFALFA Survey Sample\label{forty}}

ALFALFA catalogs have been extracted for a large contiguous region in the southern Galactic hemisphere (i.e. anti-Virgo direction) (22$^{h}$ $< \alpha <$ 03$^{h}$, 24$^{\circ} < \delta < $32$^{\circ}$), and two regions in the northern Galactic hemisphere (i.e. Virgo direction) (16$^{h}$30$^{m}$ $< \alpha <$ 07$^{h}$30$^{m}$, 4$^{\circ} < \delta < $16$^{\circ}$ and 24$^{\circ} < \delta < $ 28$^{\circ}$), with coverage totaling 2607 deg$^{2}$ or $\sim 40\%$ of the final ALFALFA volume. This includes the previously published catalogs with a total of 2706 extragalactic source measurements \citep{2009ApJS..183..214M,2009AJ....138..338S,2008AJ....136..713K,2008AJ....135..588S,2007AJ....133.2569G} in addition to an upcoming large online data release (Haynes et al. 2010, in prep)\footnote{This data release includes an additional strip of coverage,22$^{h}$ $< \alpha <$ 03$^{h}$, 14$^{\circ} < \delta < $16$^{\circ}$, which is excluded here in favor of large contiguous areas.}. This primary dataset includes both Code 1 (n = 10452) and Code 2 (n = 2750) galaxies in addition to Code 9 (n = 629) HVCs, where this figure includes measured subcomponents of larger cloud complexes.

From the primary dataset, we have selected the 40$\%$ ALFALFA Survey sample, hereafter $\alpha.40$. This sample has been selected to include only Code 1 objects, and the total sample size is further reduced by the exclusion of galaxies found beyond 15,000 \kms, where radio frequency interference from FAA radar makes ALFALFA blind to cosmic emission in a spherical shell $\sim 10$ Mpc wide. The final $\alpha.40$ sample contains 10,119 Code 1 galaxies, for a detection rate 3.9 deg$^{-2}$ compared with the HIPASS detection rate $\sim 0.2$ deg$^{-2}$ (5317 extragalactic sources over 29,000 deg$^{2}$; \citet{2004MNRAS.350.1195M} and \citet{2006MNRAS.371.1855W}). While rich in absolute number, HIPASS does not extend deep enough in redshift to sample a cosmologically fair volume.

\begin{figure}[ht!]
\epsscale{0.9}
\plotone{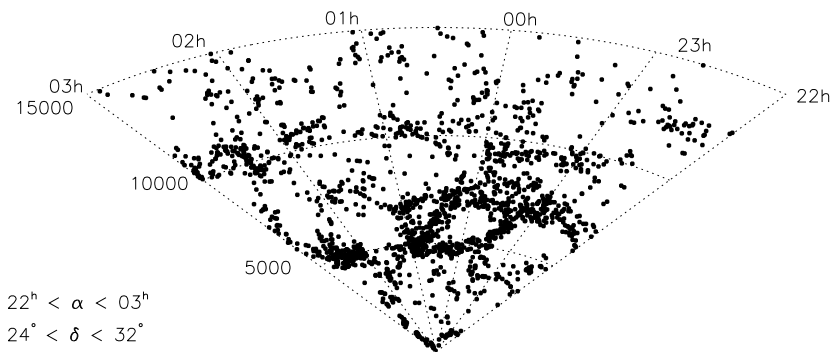}
\caption {Distribution of 2,004 sources in the $22^{h} < \alpha < 03^{h}$, $24^{\circ} < \delta < 32^{\circ}$ portion of the $\alpha.40$ sample, plotted as R.A. vs. observed heliocentric recession velocity in \kms. \label{conefall}}
\end{figure}

\begin{figure}[ht!]
\epsscale{0.9}
\plotone{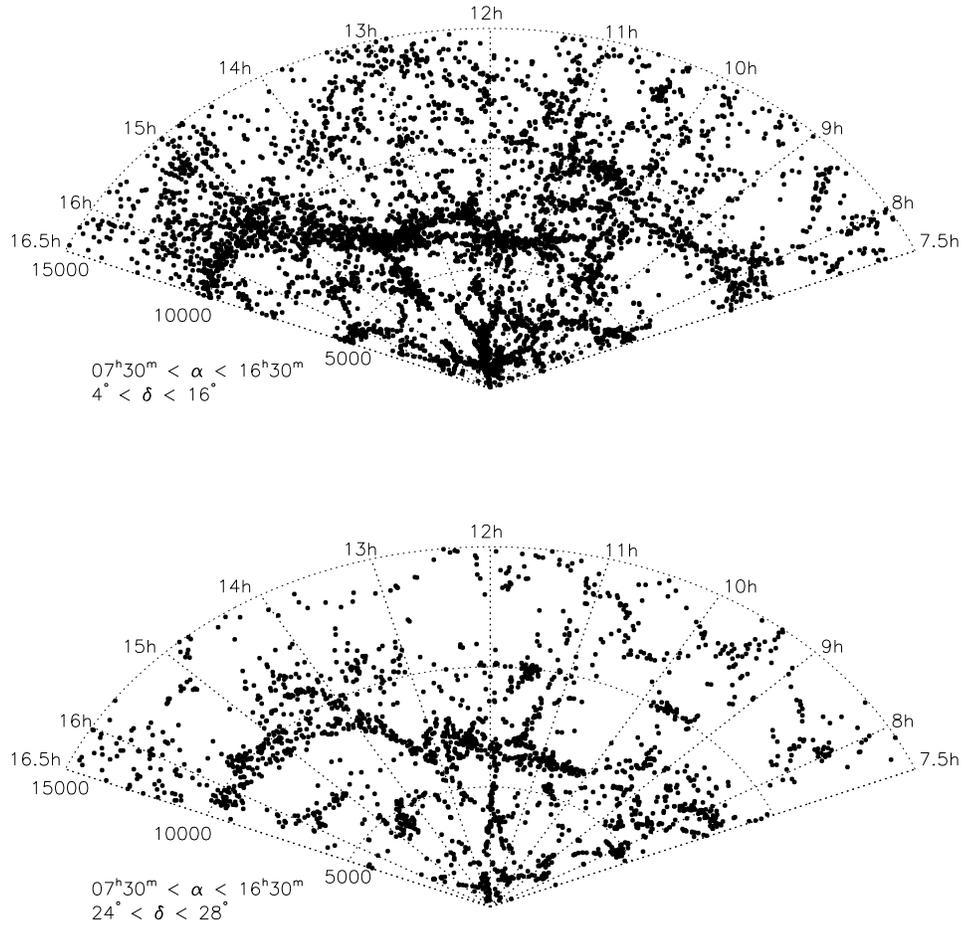}
\caption {Top panel: Distribution of 5,960 sources in the $07^{h}30^{m} < \alpha < 16^{h}30^{m}$, $4^{\circ} < \delta < 16^{\circ}$ portion of the $\alpha.40$ sample, plotted as R.A. vs. observed heliocentric recession velocity in \kms. Bottom panel: 2,155 sources over the same R.A. range as above, with $24^{\circ} < \delta < 28^{\circ}$.\label{conespring}}
\end{figure}

In Figs. \ref{conefall} and \ref{conespring} we present the redshift distribution of the 10,119 Code 1 objects in $\alpha.40$ as a set of cone diagrams by region in the survey. The two most obvious features in Fig. \ref{conefall} are the prominent void in the foreground of the Pisces-Perseus supercluster, leading to the dearth of detections out to about 3000 \kms, and the portion of the main ridge of that supercluster that cuts across the diagram. In the top panel of Fig. \ref{conespring}, the nearby Virgo cluster is prominent, as is the Coma supercluster. ALFALFA probes a wide variety of environments in the local universe, and will soon study the overall properties of HI-selected galaxies as a function of environment (Saintonge et al. 2010, in prep). 

\begin{figure}[ht!]
\epsscale{0.5}
\plotone{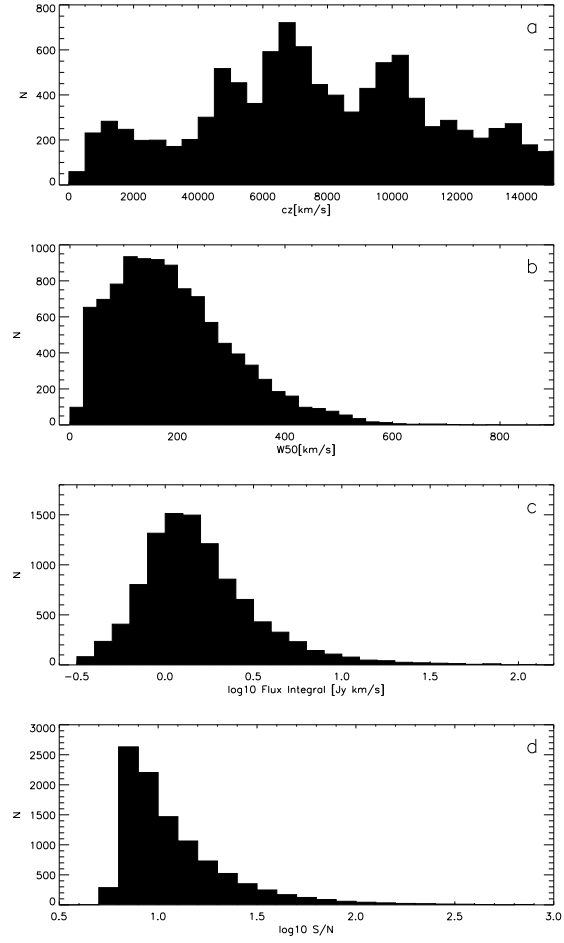}
\caption {Histograms of the galaxy properties within $\alpha.40$: (a) heliocentric recession velocity in \kms; (b) HI line width at half power (W50) in \kms; (c) logarithm of the flux integral in Jy \kms; (d) logarithm of the S/N. \label{hists}}
\end{figure}

\begin{figure}[ht!]
\epsscale{0.8}
\plotone{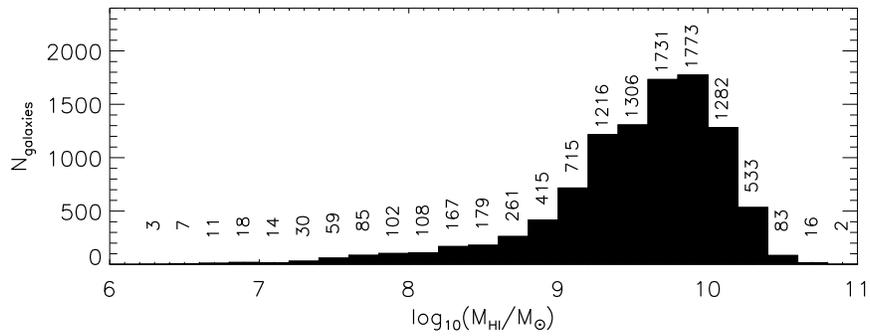}
\caption {Histogram of the distribution of HI masses in the sample, plotted as logarithm of the HI mass in solar units. \label{masshist}}
\end{figure}

Fig. \ref{hists} displays histograms of the statistical properties of the $\alpha.40$ sample. From (a) to (d), these histograms represent the heliocentric velocity, velocity width W$_{50}$, integrated flux S$_{int}$, and S/N properties. In particular, note that the S/N is high for all detections, since Code 2 objects have been excluded from this analysis. For clarity, the histogram of the HI masses of galaxies in the sample is plotted separately, in Fig. \ref{masshist}. On the low mass end, where ALFALFA can place strong constraints on the faint-end slope of the HIMF, the $\alpha.40$ sample contains $\sim$ 340 galaxies with $\log (M_{HI}/M_{\odot}) < 8.0$ and $\sim$ 114 with $\log (M_{HI}/M_{\odot}) < 7.5$; on the high mass end, which is best probed by surveys with deep redshift limits, there are $\sim$ 50 galaxies with $\log (M_{HI}/M_{\odot}) > 10.5$.

The large sample size of ALFALFA, extending over a range of HI masses, is one of its key strengths in relation to the problem of characterizing the density of neutral gas in the present-day universe. With such a large number of galaxies, we can approach our calculation of the HIMF in two distinct ways. First, using the entire sample and a well-known characterization of our sensitivity, we can apply corrections and obtain the overall function without excluding sources. Second, however, we can make stringent integrated flux cuts and use only those galaxies bright enough to be detectable irrespective of other properties (e.g. profile width). The sample contains $\sim$ 3500 galaxies with an integrated flux $> 1.8$ Jy \kms, which provides a strict cut above which our objects are detected regardless of profile width. This subsample size is comparable to the full sample size for previously-published HIMFs such as HIPASS, but samples a fair cosmological volume. This subsample, referred to hereafter as $\alpha.40_{1.8}$, provides a test case for analyzing the quality of the HIMF measurement for the full $\alpha.40$ sample. The precise details of the calculation, of ALFALFA's sensitivity, and of the corrections applied to the HIMF calculated from $\alpha.40$, make up the bulk of the following sections and of Appendices A and B.

\section{Determination of the HIMF\label{method}}
\subsection{The HI Mass Function \label{himf}}

The HI mass function, like galaxy luminosity functions, is usually parametrized as a Schechter function of the form
\begin{equation}
\phi(M_{HI}) = \frac{dn}{d \log M_{HI}} = \ln 10 \ \phi_{*} \ \left (\frac{M_{HI}}{M_{*}} \right )^{\alpha + 1} \ e^{-\frac{M_{HI}}{M_{*}}}
\label{eqnschech}
\end{equation}

\noindent The parameters of interest are the faint-end slope $\alpha$, the characteristic mass $\log M_{*}$, and the scaling factor $\phi_{*}$.

$\phi(M_{HI})$ has historically been calculated in one of two ways. The $\Sigma 1/V_{max}$ method \citep{1968ApJ...151..393S} can be understood by analogy to a purely volume-limited sample, in which case the HIMF would be obtained by the galaxy counts divided by the total volume of the survey. The $\Sigma 1/V_{max}$ method treats each individual galaxy in this way, by weighting the galaxy counts by the maximum volume $V_{max,i}$ within which a given source could have been detected. This weighting strategy allows the inclusion of low-mass galaxies, visible only in the nearby Universe, in the same sample as rare high-mass galaxies, found only in larger volumes. Additionally, the weights may be adjusted in order to correct for a variety of selection effects, large-scale structure effects, and missing volume within the dataset, so that a well-characterized survey can robustly measure the HIMF.

An alternative method, the Two Dimensional Step-Wise Maximum Likelihood (2DSWML) approach, was applied to the HIPASS measurements of the HIMF \citep{2003AJ....125.2842Z,2005MNRAS.359L..30Z}. This method is designed to make the calculation of the HIMF less sensitive to local large scale structure, since shallow blind HI catalogs are contaminated by the richness of the Local Supercluster. If $1/V_{max}$ is used without correction for this overdensity, the resulting HIMF will overestimate the contribution by low-mass galaxies and steepen the faint-end slope $\alpha$. Stepwise maximum likelihood methods, by contrast, are designed to reduce this bias, by assuming that the shape of the HIMF is the same everywhere and then obtaining the $\phi (M_{HI})$ that maximizes the probability of the observed distribution \citep{1988MNRAS.232..431E}. Given the dependence of the ALFALFA survey's sensitivity on both mass and profile width (\S \ref{width}), a Two-Dimensional Step-wise Maximum Likelihood (2DSWML) approach is necessary to calculate the HIMF for the full sample \citep{2000MNRAS.312..557L}. 2DSWML maintains the main advantages of the SWML method, which are its robustness against density fluctuations in the survey volume and its model-independent approach.

In this work, we apply both the 1/V$_{max}$ and the 2DSWML method for various reasons. Given our knowledge of our sample's characteristics and sensitivity, the 1/V$_{max}$ method is simple to apply and straightforward to assess for potential bias. We can account for large scale structure and other selection effects by applying well-motivated corrections (discussed in \S \ref{corrs}). Perhaps more significantly, this method also allows us to quantify and understand those effects on the ALFALFA survey. In particular, a goal of ALFALFA is to further probe the differences between HI mass functions in different environments; the 2DSWML assumption that the shape of the function is the same throughout a sample may not be valid. By contrast, the 2DSWML method is designed to be more resistant to effects from large-scale structure, and also results in a calculation of the selection function which can be used in future analysis of the sample via, for example, the two-point correlation function. A comparison of the $1/V_{max}$ and 2DSWML methods as applied to $\alpha.40$ is considered in \S6 .

In both the 1/V$_{max}$ and 2DSWML analyses, we have used 5 mass bins per dex, and have found that the HIMF is not strongly affected by choice of bin size. In the case of 2DSWML, we also bin by profile velocity width, and find no significant difference for bin sizes between 2 and 20 bins per dex. The two main sources of error are counting statistics within the bins and mass errors.

\subsection{Errors on Distances and Masses\label{disterrs}}

Minimizing and taking into account distance errors is key to robust estimation of luminosity and mass functions, in particular at the faint end. \citet{2004ApJ...607L.115M} considered how strongly distance uncertainties will tend to affect a given local volume survey's estimate of the faint-end slope of the mass function. In that work, the authors accounted for distance errors by constructing a mock catalog, with masses assigned from a chosen HIMF and with the spatial distribution determined from the density field of the IRAS Point Source Catalog Redshift survey (PSCz; \citet{1999MNRAS.308....1B}). They concluded that a survey toward the Virgo cluster, like a portion of the sample considered here, will overestimate distances to those galaxies if pure Hubble flow is used, since objects in that field are falling into Virgo. Since the HI mass depends on distance as D$^{2}$, this has serious consequences for the faint-end slope of the HIMF. Therefore, work in this region relies both on the development of well-constrained local velocity models from primary and secondary distance catalogs and on a careful consideration of the effects of distance uncertainties. We consider the Virgo cluster as a special case of this general problem in \S \ref{excludevirgo}.

These difficulties arise precisely because the lowest mass objects can be detected only at small distances, so that the fractional distance errors due to deviations from Hubble flow most strongly affect the most interesting bins of the mass function. The best distance estimates, primary distances based on, e.g., Cepheids or the tip of the red giant branch, can only estimate distances to within $\sim$ 10$\%$ error, so beyond cz $\sim$ 6000 \kms \ the uncertainties on distances obtained via a primary method and those obtained assuming pure Hubble flow become comparable, and the latter is typically used for simplicity. Within that distance, however, the distance uncertainties can have a very strong influence, up to 100$\%$ in the case of the Virgo cluster.

To minimize distance uncertainties, the ALFALFA survey has adopted a distance estimation scheme that makes use of a peculiar velocity flow model for the local Universe \citep{2005PhDT.........2M}. This parametric multiattractor model, based on the SFI++ catalog of galaxies with Tully-Fisher distances \citep{2007ApJS..172..599S}, includes two attractors (Virgo and a Great Attractor) along with a dipole and quadrupole component. Distances to almost all $\alpha.40$ galaxies within 6000 \kms \ are estimated from the flow model. Beyond cz$_{CMB}$ = 6000 \kms, the model is not well-constrained, so distances are estimated from Hubble flow (H$_{0}$ = 70 km $s^{-1}$ Mpc$^{-1})$. Within 6000 \kms, some galaxies have measured primary distances, which are applied in our scheme, and other galaxies are known to belong to a group, in which case the group's mean velocity is used for distance estimation. The \citet{2005PhDT.........2M} flow model also provides error estimates, constrained by the fit of the model to the observed velocity field and with a minimal error based on the local velocity dispersion 163 \kms. When distances are estimated using pure Hubble flow, the error is estimated to be $\sim 10\%$ via the assumption that peculiar velocities are $\sim$ a few hundred \kms.

Mass errors for individual galaxies in our sample are calculated from the measured error on the integrated flux and an estimated error on the distance, which is the larger of the local velocity dispersion 163 \kms, the distance error estimate of the \citet{2005PhDT.........2M} flow model, or 10$\%$ of the distance. Because the mass error shifts galaxies into different bins of the HIMF, the relationship between these errors and the final HIMF parameter errors is complex. We deal with these errors by calculating several hundred realizations of the HIMF after randomly assigning flux and distance errors to each galaxy to find the spread in each mass bin. 

There is a complication on the high-mass end of the sample, as well. Arecibo's relatively large beam size at 21 cm ($\sim$ 3.5 arcmin) can cause source confusion at large distances, where we also find our largest-mass objects. When this occurs, ALFALFA may be detecting more than one individual gas-rich galaxy as a single source, but in cases of interaction it's also possible that the galaxies involved are part of a single, large HI envelope. While higher-resolution followup would be required to fully resolve this issue, we have investigated optical images and redshift catalogs for the highest mass ($\log M_{HI} >$ 10.5) ALFALFA detections, and have found that the majority of these objects are not likely to be blends of HI emission from an interacting system and some others are close pairs that are likely to share a single gas envelope.

\subsection{1/V$_{max}$ Method \label{corrs}}

For each galaxy in $\alpha.40$, $V_{max,i}$ is calculated based on that galaxy's HI mass M$_{i}$, the minimum integrated flux S$_{min,i}$ at which such a galaxy is detected in ALFALFA, and finally the distance D$_{max,i}$ corresponding to that limit. The calculated $V_{max,i}$, corresponding to the effective search volume for that galaxy, excludes volume that is not covered by ALFALFA, including volumes where detection ability, and therefore effective search volume, is reduced by the appearance of radio frequency interference at the corresponding frequency. Galaxies are binned by mass and $\phi (M_{HI})$ is calculated by summing the reciprocals of $V_{max}$. 

By weighting the count for each galaxy, the $1/V_{max}$ method can be corrected for a variety of known systematic effects. The major corrections applied to the HIMF for this sample address (1) missing volume, (2) the profile width-dependent sensitivity of the survey, and (3) the known large scale structure in the local volume. 

\begin{figure}[ht!]
\epsscale{0.6}
\plotone{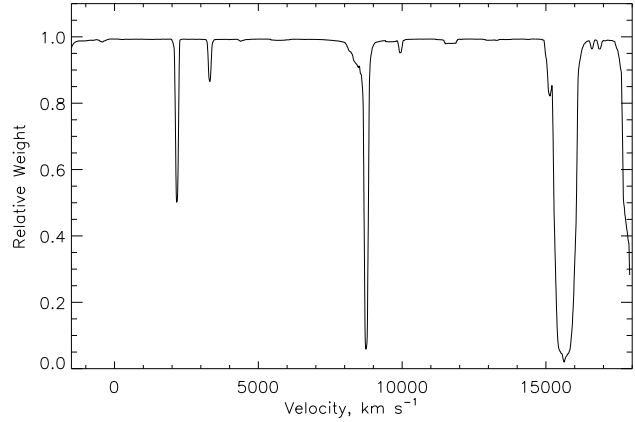}
\caption {The average relative weight within the 40$\%$ ALFALFA survey volume as a function of observed heliocentric velocity. Where the relative weight is near 1.0, nearly the entire surveyed volume was accessible for source extraction, and the regions of lower relative weight correspond to manmade radio frequency interference. These sources are not always present, and do not always result in a complete loss of signal, so there are regions where the average weight is reduced only modestly.The large dip between 15000 and 16000 \kms \ is due to the FAA radar at the San Juan airport, and because of this extreme loss of volume at large distances we restrict our sample to only those galaxies within 15000 \kms. \label{avweight}}
\end{figure}

Sources of radio frequency interference contaminate the signal in regions of frequency space corresponding to spherical shells in the survey volume. This effectively reduces the search volume of the overall survey. Fig. \ref{avweight} shows the average relative weight, compared to 100$\%$ coverage, within the $\alpha.40$ survey volume as a function of velocity. The large dip between 15000 and 16000 \kms \ is due to the FAA radar at the San Juan airport, and because of this extreme loss of volume at large distances we restrict the $\alpha.40$ sample to only those galaxies within 15000 \kms. Given our knowledge that the relative weight is less than 1.0 at specific distances, the V$_{max}$ value calculated for a specific galaxy is reduced to reflect the loss of effective search volume. This correction is not significant for the lowest-mass galaxies, but more generally, the correction is very small. The effect on the final Schechter parameters for the HIMF is on the order of 2$\%$.

\begin{figure}[ht!]
\epsscale{0.7}
\plotone{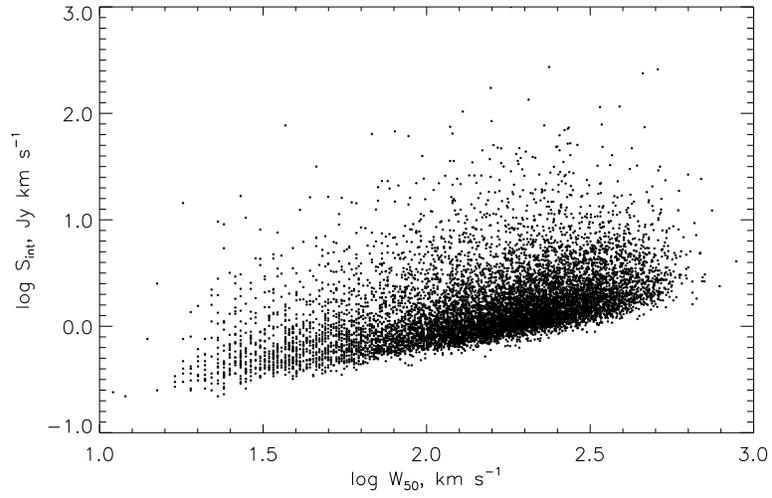}
\caption {The distribution of sources detectable by ALFALFA, which is dependent on both flux S$_{int}$ in Jy \kms \ and profile width W$_{50}$ in \kms. \label{sensitivity}}
\end{figure}

As discussed in \ref{width}, ALFALFA's detection ability is dependent on the profile velocity width of the signal, W$_{50}$, in \kms, rather than strictly on the integrated flux of the signal. To obtain an expression for this detection limit, we used the data itself, as displayed in Fig. \ref{sensitivity}. The dependence of ALFALFA sensitivity on both flux and profile width, described in \S \ref{width}, has the further complication of affecting the survey's completeness, and this must be accounted for in order to extract the underlying HIMF. The distribution in Fig. \ref{sensitivity} indicates that ALFALFA finds many galaxies with low fluxes and narrow widths, but there is a deficiency of galaxies with low fluxes and large widths. Because we have no knowledge of the true distribution below ALFALFA's detection capability, we have developed a completeness correction that takes advantage only of the data, making no assumptions about the potentially intrinsically small unobserved population. The profile width completeness correction most strongly affects galaxies with $\sim 9.0 < \log (M_{HI}/M_{\odot}) < 10.0$, and has a very small influence ($< 2\%$) on both the faint-end slope $\alpha$, since low-mass (i.e. narrow velocity width) galaxies aren't affected, and $\log (M_{*})$, since the counts in the high-mass bins are large enough to robustly constrain this. This is essentially a galaxy counting correction, so its primary influence is on $\phi_{*}$, increasing that parameter by a factor of 20$\%$. The full details of this completeness correction are described in Appendix A. The validity of this completeness correction, which we have applied to the full sample, is tested in \S \ref{global} and \ref{swmlglobal} by calculating the HIMF using an integrated flux cut, which allows us to neglect the biased sensitivity dependence on width. By comparing the resulting HIMF in both cases, we assess the impact of this correction.

The most significant bias in the 1/V$_{max}$ calculation of the HIMF is that due to the large-scale structure of the galaxy distribution. Blind HI surveys tend to be relatively shallow and are thus biased by the overdensity of the local volume, which particularly affects the lowest-mass HI-rich galaxies that are only found nearby. If a correction for large-scale structure is not applied, we overestimate the impact of low-mass galaxies on the overall HIMF, therefore boosting the faint-end slope $\alpha$ artificially. We discuss this correction in Appendix A. The large-scale structure volume correction has only a very weak effect on $\log (M_{*})$, but the effects on $\alpha$ ($\sim 10\%$) and $\phi_{*}$ ($\sim 30\%$) are large. Since this correction is so significant, it is sensitive to the details of the density reconstruction used. Agreement between the 1/V$_{max}$ and 2DSWML results provide the best indication of the quality of this correction.

However, large-scale structure introduces the further bias of selectively reducing counts in mass bins that are primarily detectable in void volumes, and the weighting scheme correction cannot account for that. The voids in the Pisces-Perseus region between 3000 and 8000 \kms, visible in Fig. \ref{conefall}, in particular, bias that portion of the $\alpha.40$ sample against galaxies with $8.5 < \log (M_{HI}/M_{\odot}) < 9.0$, leading to a systematic undercounting in those bins. Because the 1/$V_{max}$ method is sensitive to large scale structure, this undercounting introduces a spurious `bump' feature into the HIMF, describe in detail in Section \ref{global}.

\subsection{2DSWML Method\label{swml}}

As discussed in \S \ref{himf} and \ref{corrs}, the main disadvantage of the $1/V_{max}$ method is its potential sensitivity to large-scale structure. If large-scale structure corrections were not adopted, the density of low HI-mass galaxies would be systematically overestimated, since most of these galaxies are detectable only in the very local, substantially overdense universe, including the Virgo Cluster and the Local Supercluster. This would bias the low-mass slope of the Schechter fit to the HIMF ($\alpha$), weakening one of the major strengths of the ALFALFA dataset, which is its ability to probe the population of extremely low HI-mass galaxies over a wide solid angle for the first time.

The original SWML method is applicable to samples selected by integrated flux. It assumes that the observed galaxy sample is drawn from a common HI mass function throughout the survey volume, denoted by $\phi(M_{HI})$. Unlike most Maximum Likelihood methods, which assume a functional form for $\phi (M_{HI})$ \citep{1979ApJ...232..352S}, SWML splits the distribution in bins of $m = \log(M_{HI}/M_\odot)$ and assumes a constant distribution within each logarithmic bin. In this way, the value of the distribution in each of the bins becomes a parameter, $\phi_j$ ($j =1,2,...,N_m$), which is adjusted in order to maximize the joint likelihood of detecting all galaxies in the sample, hence yielding a Maximum Likelihood estimate of the mass distribution. Since the values of the parameters are free to vary independently, the procedure above is completely general and does not assume any functional form for the distribution \textit{a priori}.

In the case where the sample is not integrated-flux$-$limited and the selection function depends on additional observables, the SWML technique has to be extended to take into account the underlying galaxy distribution in all the physical properties that enter the calculation of the selection function. In the case of $\alpha$.40, the limiting integrated flux depends on the galaxy profile width W$_{50}$ and thus the method needs to consider the joint two-dimensional distribution of galaxies in both HI mass and observed velocity width, $\phi(M_{HI},W_{50})$. 2DSWML relies on the assumption that the sample is statistically complete. Since ALFALFA's sensitivity to a source is dependent on both its integrated flux and its velocity width W$_{50}$ (\S \ref{width}), we fit a strict completeness threshold to the observed relationship as seen in Fig. \ref{sensitivity} and exclude galaxies falling below this completeness cut.

The details of the 2DSWML method and its application to $\alpha$.40 are given in Appendix B.

\subsection{HIMF Error Analysis\label{errors}}

The simplest source of error in the estimate of the HIMF is from Poisson counting errors in the bins, which is added to the other sources of error considered next. The relationship between errors on corrections applied to individual galaxies and errors on the final HIMF points and measured parameters is complex. Mass errors, for example, may shift galaxies in the sample from one bin to another as discussed in \S \ref{disterrs}, so it is not possible to analytically calculate the error on a particular bin. In order to treat these errors appropriately, we create $> 250$ realizations of the HIMF for each of the results shown in \S4 and 5. The error on S$_{int}$ is measured in the ALFALFA source extraction pipeline, and we have estimated errors on the distance for each galaxy in the sample. Each of these contributes to the mass error, and we apply a Gaussian random error to each galaxy's mass in each realization. The spread in the bin values across the ensemble of realizations contributes to the overall error in each point. We consider errors due to uncertain parameter estimation in the relationship between $\log (M_{HI}/M_{\odot})$ and the Gumbel distribution parameters $\mu$ and $\beta$ in the same way. This results in a HIMF that has taken known sources of error into consideration.

Sources of systematic bias remain, particularly for the 1/V$_{max}$ measurement which is sensitive to the large-scale structure in the galaxy distribution. The effects of large-scale structure and of cosmic variance will be reduced as the survey continues, increasing its volume and coverage of varied cosmological environments.

In order to account for errors that are more difficult to quantify, we follow the example of \citet{2005MNRAS.359L..30Z} and jackknife resample 21 equal-area regions. The resampling technique will help account for residual large-scale structure beyond that which we have corrected, and also for any systematic survey effects that change spatially across the sky or temporally throughout the survey's observations.

\subsubsection{2DSWML Error Estimates \label{swmlerrorsection}}
The 2DSWML approach introduces another source of error. We assign errors on the parameters $\phi_{jk}$, introduced in \S \ref{swml}, via the inverse of the information matrix following \citet{2000MNRAS.312..557L} and \citet{1988MNRAS.232..431E}. The general form of the information matrix for a likelihood function $\mathcal{L}$ that depends on a set of parameters $\mathbf{\theta}$ is given by

\begin{eqnarray}
\mathbf{I}(\mathbf{\theta}) = 
 - \left[ \begin{array}{cc}
\frac{\partial^2}{\partial \theta_m \partial \theta_n}\ln \mathcal{L}+\frac{\partial}{\partial \theta_m} g \frac{\partial}{\partial \theta_n}g & \frac{\partial}{\partial \theta_n}g \\
 \frac{\partial}{\partial \theta_m}g& 0 \end{array} \right]   \label{infotemplate}
\end{eqnarray}   

\noindent
where $g$ is a constraint of the form $g(\mathbf{\theta})=0$. We choose to apply the constraint $g=\sum_j \sum_k \: (\frac{M_{HI,j}}{M_{HI,ref}})^{\beta_1}(\frac{W_{50,k}}{W_{50,ref}})^{\beta_2} \: \phi_{jk} \: \Delta m \Delta w-1=0$, with $\beta_1 = \beta_2 =1$ and reference values for the HI mass and W$_{50}$ equal to the $\alpha.40$ sample mean. The result is an error estimate for the parameters $\phi_{jk}$, i.e. the value of the HIMF in each mass bin, and is added in quadrature to the other sources of error described above.   

\section{1/V$_{max}$ Method: Results \label{results}}

\subsection{Global HI Mass Function and $\Omega_{HI}$ \label{global}}

\begin{figure}[ht!]
\epsscale{0.8}
\plotone{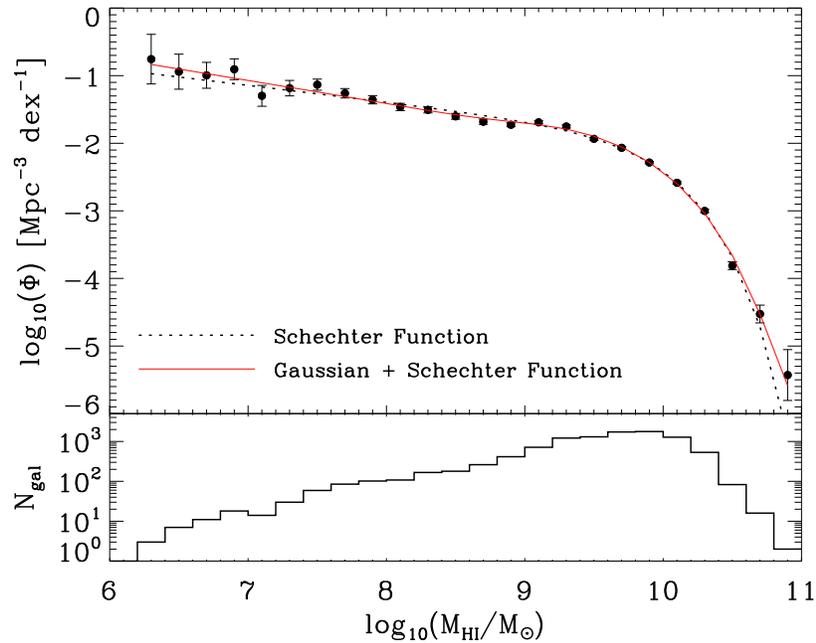}
\caption {The global HI mass function derived from $\alpha.40$ via the 1/V$_{max}$ method. Points are the HIMF value, per dex, in each mass bin, with errors as described in the text overplotted. The black dotted line is the Schechter function fit to the points, and the red solid line is the sum of a Schechter function and a Gaussian fit to the points. The histogram, bottom panel, shows the logarithm of the bin counts.\label{globalHIMF}}
\end{figure}

The global HI mass function derived from the $\alpha.40$ sample via the 1/V$_{max}$ method is presented in the top panel of Fig. \ref{globalHIMF}. Overplotted error bars have been derived as described above; mass errors due to errors on flux and distance estimates are reflected in the errors on the HIMF points, rather than on the mass-axis bin positions, since these errors change the bin counts. 

The best-fit Schechter function describing this HIMF is overplotted as a dashed line. The derived parameters are $\phi_{*}$ (h$_{70}^3$ Mpc$^{-3}$dex$^{-1}$) = 6.0 $\pm$ .3 $\times$ 10$^{-3}$, $\log $(M$_{*}/$M$_{\odot})$ + 2 $\log$ h$_{70}$  = 9.91 $\pm$ 0.01 and $\alpha$ = -1.25 $\pm$ 0.02. However, the large-scale structure in the ALFALFA survey regions has introduced a `bump' into this measurement of the HIMF. The feature visible in Fig. \ref{globalHIMF} at $\log (M_{HI}/M_{\odot}) \sim 9.0$ does not appear to be intrinsic to the HI-rich galaxy population. However, previous work on luminous galaxies has suggested that the shape of luminosity and mass functions may be more complex than single Schechter functions. Luminosity functions in clusters, such as Coma and Fornax, are inconsistent with single values of $\alpha$; \citet{1998MNRAS.294..193T} has recommended a `composite' luminosity function that steepens for both bright and faint objects and flattens out in between, which provides a `dip' feature. Single Schechter functions provided a poor fit to 2dFGRS luminosity functions \citep{2002MNRAS.333..133M}, and results from the Sloan Digital Sky Survey also suggest that a second \citep{2004ApJ...600..681B} or third \citep{2009MNRAS.398.2177L} Schechter function component best describes the underlying population of galaxies at low redshift. While, given these findings, it is possible that the feature in Fig. \ref{globalHIMF} suggests a complex shape in the HIMF, it is more likely that the feature is spurious, as we discuss below.

Such features occur because the 1/V$_{max}$ method is sensitive to large scale structure. Because the survey's HI mass sensitivity varies with distance (i.e., $\alpha$.40 is not a volume-limited sample), each mass bin in the HIMF corresponds to some preferred distance at which ALFALFA is most sensitive to galaxies in that mass bin. Extended large-scale structures can therefore change the shape of the HIMF in bins corresponding to the distance of those features. Because of the large sample size of $\alpha$.40, it is possible to separately investigate the three survey regions represented by the cone diagrams in Figs. \ref{conefall} and \ref{conespring} and to isolate the structures that contribute to such features. Specifically, the `bump' feature in Fig. \ref{globalHIMF} is due to a lack of sources in the foreground of the Great Wall and an overabundance within the Great Wall, clearly evident in Fig. \ref{conespring}. The large scale structure correction (\S \ref{corrs} and Appendix A) reduces this feature, but cannot totally eliminate it, in part because density maps used to correct for large scale structure are smoothed to $\sim$ a few Mpc scales and can underestimate extremes in the density contrast. Features such as this one will be reduced as the ALFALFA survey continues and the sample grows. The 2DSWML method is not sensitive to large scale structure and does not produce this feature (\S5).

\begin{figure}[ht!]
\epsscale{0.5}
\plotone{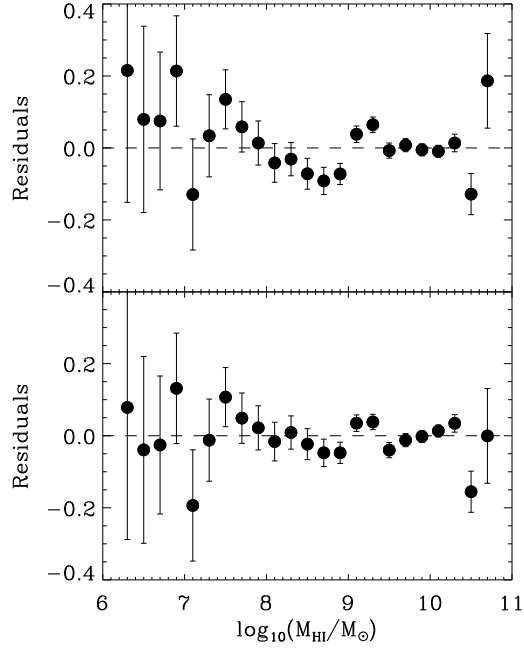}
\caption {The residuals between the 1/V$_{max}$ HIMF points and the derived best-fit Schechter function (top panel) and the best-fit sum of a Schechter and a Gaussian (bottom panel). Bars represent the errors on each point, to show the significance of the residual in each case. The Schechter function provides a poor fit to the spurious `bump' feature, and this effect is reduced by the addition of a Gaussian component. The highest-mass bin, which has a large error value, is excluded from this plot.\label{HIMFresid}}
\end{figure}

This feature appears significant in part because our statistical errors on the HIMF points are so small, but it leads to a poor fit and an underestimate for the faint-end slope $\alpha$. This is clear in Fig. \ref{HIMFresid}, which displays the residual between the 1/V$_{max}$ HIMF points and the derived best-fit Schechter function in the top panel and shows that the Schechter function systematically over- and under-estimates the HI mass function due to this feature.

While this feature is well-understood, it has the undesirable effect of artificially reducing the faint-end slope $\alpha$. In an effort to reduce the effect of this spurious feature and to better fit the points, we fit the sum of a Schechter function and a Gaussian; the Gaussian component serves to filter out the feature, leading to a better estimate of $\alpha$. The results are shown as the solid line in Fig. \ref{globalHIMF} with the residuals shown in the bottom panel of Fig. \ref{HIMFresid}. This fit significantly improves the reduced $\chi^{2}$, and the residuals are small and, near $\log (M_{HI}/M_{\odot}) \sim 9.0$, more randomly scattered about 0 in contrast to the top panel of Fig. \ref{HIMFresid}. However, there is larger uncertainty in the parameters in this case, since each function is constrained by fewer points. The Schechter function parameters, displayed in Table \ref{params}, are $\log $(M$_{*}/$M$_{\odot})$ + 2 $\log$ h$_{70}$  = 9.95 $\pm$ 0.04 and $\alpha$ = -1.33 $\pm$ 0.03. The Schechter function measurement of $\phi_{*}$ (h$_{70}^3$ Mpc$^{-3}$dex$^{-1}$) = 3.7 $\pm$ .6 $\times$ 10$^{-3}$, however, has been affected by the addition of the second component to the fit, and we therefore defer to the 2DSWML measurement of that parameter.

The Gaussian parameters are not included in Table \ref{params}, since they are used to filter out the `bump' feature and are not expected to have physical meaning. The best-fit Gaussian has peak height (h$_{70}$ Mpc$^{-3})$ 5 $\pm$ 1 $\times$ 10$^{-3}$, mean $\log $(M$_{\mu}/$M$_{\odot})$ + 2 $\log$ h$_{70}$ 9.28 $\pm$ 0.06 and spread in $\log $(M$_{\mu}/$M$_{\odot})$ + 2 $\log$ $\sigma$ = 0.41 $\pm$ 0.03. 

We conclude that the proper values of $\alpha$ and $\log $(M$_{*}/$M$_{\odot})$ extracted from the 1/V$_{max}$ method are -1.33 $\pm$ 0.03 and 9.95 $\pm$ 0.04, respectively. Table \ref{params} lists both the spurious 1/V$_{max}$ Schechter function parameters as well as the parameters found when a Gaussian is added to fit the spurious feature. The addition of the Gaussian brings the 1/V$_{max}$ results for the parameters $\alpha$ and M$_{*}$ into excellent agreement with the 2DSWML method and the flux-limited $\alpha.40_{1.8}$ subsample results. 

As an additional test of our corrections for profile width sensitivity, we have derived the 1/V$_{max}$ HIMF from the integrated flux-limited subsample $\alpha.40_{1.8}$ (described in \S \ref{forty}). This mass function is corrected for large-scale structure and include mass errors, but is not subject to the same bias against broad HI profiles. The $\alpha.40_{1.8}$ HIMF is well-fit by a pure Schechter function. The results are listed in Table \ref{params}. The $\alpha.40_{1.8}$ HIMF is consistent with those derived from the full $\alpha.40$ sample. We therefore conclude that our survey sensitivity is well-characterized and that our measurements based on the full sample are complete and representative. However, since this limited sample does not probe the galaxies at the extremes of the mass function, it is subject to larger errors on the points and in the parameters. 

\subsubsection{Measurement of $\Omega_{HI}$ \label{omegasec}}
The density $\Omega_{HI}$ of neutral hydrogen in the local Universe, expressed in units of the critical density, can be calculated in two ways from the derived HI mass function. Integrating analytically over the best fit Schechter function gives $\Omega_{HI} = \phi_{*} \, M_{*} \, \Gamma(2+\alpha)$= 4.4 $\pm$ 0.3 $\times 10^{-4}$ h$^{-1}_{70}$, slightly (16$\%$) higher than the final HIPASS value 3.7 $\times 10^{-4}$ h$^{-1}_{70}$ \citep{2005MNRAS.359L..30Z}. Using the binned points directly, we find the same result: $\Omega_{HI}= $ 4.4 $\pm$ 0.1 $\times 10^{-4}$ h$^{-1}_{70}$. This agreement is an indication that our findings are well-represented in the high-mass bins by our Schechter function fit, despite the spurious feature. $\Omega_{HI}$ carries a small error since it is negligibly affected by the mass and distance errors on the faint end.

\begin{figure}[ht!]
\epsscale{0.5}
\plotone{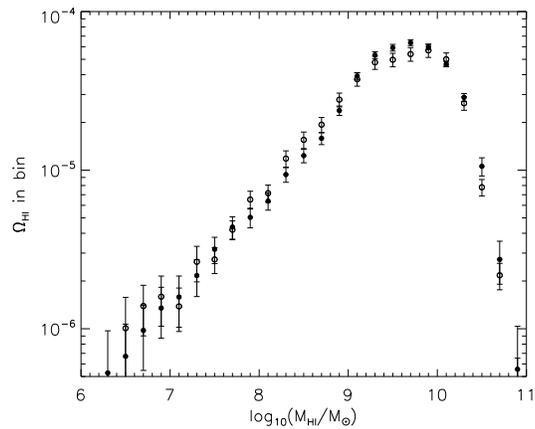}
\caption {The contribution to $\Omega_{HI}$ by the galaxies in each bin in $\alpha.40$. Filled circles have been calculated via the 1/V$_{max}$ method, and open circles are from the 2DSWML method. The total density of neutral hydrogen in the local Universe is dominated by galaxies with $9.0 < \log (M_{HI}/M_{\odot}) < 10.0$.\label{omega}}
\end{figure}

In Fig. \ref{omega}, we show the contribution of each 1/V$_{max}$ mass bin to $\Omega_{HI}$ as filled circles. The total density of neutral hydrogen in the local Universe is dominated by galaxies with $9.0 < \log (M_{HI}/M_{\odot}) < 10.0$, and in these bins we measure the HIMF to be larger than \citet{2005MNRAS.359L..30Z} do, thus finding a larger value of $\Omega_{HI}$. The ALFALFA survey extends further in redshift than HIPASS, with a median redshift $\sim$ 8000 \kms \ compared to $\sim$ 3000 \kms, allowing us to detect significantly more high-mass objects (\S \ref{prevresult}).

\section{2DSWML Method: Results \label{swmlresults}}

\subsection{Global HI Mass Function and $\Omega_{HI}$ \label{swmlglobal}}

\begin{figure}[ht!]
\epsscale{0.8}
\plotone{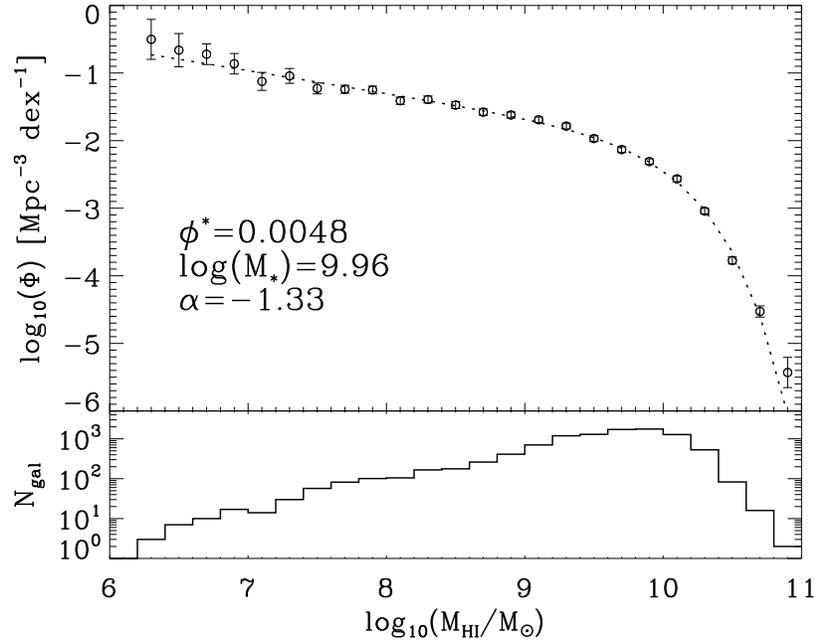}
\caption {The global HI mass function derived from $\alpha.40$ via the 2DSWML method. As in Fig. \ref{globalHIMF}, points are the HIMF value, per dex, in each mass bin, with errors as described in the text overplotted. The dotted line is the Schechter function fit to the points and the Schechter function parameters are listed. The histogram, bottom panel, shows the logarithm of the bin counts.\label{globalHIMFSWML}}
\end{figure}

The HIMF derived from $\alpha.40$ through the 2DSWML method is shown in Fig. \ref{globalHIMFSWML}. The derived parameters are $\phi_{*}$ (h$_{70}^3$ Mpc$^{-3}$dex$^{-1}$) = 4.8 $\pm$ .3 $\times$ 10$^{-3}$, $\log $(M$_{*}/$M$_{\odot})$ + 2 $\log$ h$_{70}$  = 9.96 $\pm$ 0.02 and $\alpha$ = -1.33 $\pm$ 0.02. To test the robustness of this HIMF estimate, we also applied a one-dimensional SWML approach to the flux-limited $\alpha.40_{1.8}$ sample, and found results consistent with the global, two-dimensional result ( $\phi_{*}$ = 4.5 $\pm$ .9 $\times$ 10$^{-3}$, $\log (M_{*})$  = 9.96 $\pm$ 0.04 and $\alpha$ = -1.36 $\pm$ 0.06).

\subsubsection{Measurement of $\Omega_{HI}$ \label{swmlomegasec}}

As in the case of the 1/V$_{max}$ method, we calculate the neutral hydrogen density $\Omega_{HI}$ from an analytical integration of the best-fit Schechter function and from a summation over the points themselves. From the Schechter function we find $\Omega_{HI} =$ 4.3 $\pm$ 0.3 $\times 10^{-4}$ h$^{-1}_{70}$ and from the binned points we find 4.4 $\pm$ 0.1 $\times 10^{-4}$ h$^{-1}_{70}$. In both cases, our result is consistent with the 1/V$_{max}$ method and is slightly higher than the HIPASS result. The contribution by each bin is shown in Fig. \ref{omega} as open circles.

\begin{deluxetable*}{cccccc}
\tablewidth{0pt}
\tabletypesize{\tiny}
\tablecaption{HI Mass Function Fit Parameters\label{params}}
\tablehead{ \colhead{Sample and} & \colhead{$\alpha$} & \colhead{$\phi_{*}$} & \colhead{$\log $ (M$_{*}$/M$_{\odot}$)} & \colhead{$\Omega_{HI}$, fit} & \colhead{$\Omega_{HI}$, points} \\
Fitting Function& & (10$^{-3}$ h$_{70}^{3}$ Mpc$^{-3}$ dex$^{-1}$) & + 2 $\log$ h$_{70}$ & ($\times$ 10$^{-4}$ h$^{-1}_{70}$) & ( $\times$ 10$^{-4}$ h$^{-1}_{70}$)
}
\startdata
1/V$_{max}$&-1.25 (0.02)& 6.0 (0.3) & 9.91 (0.01) & 4.4 (0.2) & 4.4 (0.1) \\
Schechter + Gaussian\tablenotemark{a} & -1.33 (0.03) & 3.7 (0.6)\tablenotemark{b}& 9.95 (0.04) & & \\
\\
1/V$_{max}$, Non-Virgo & -1.20 (0.02)& 6.1 (0.3) & 9.90 (0.01) & 4.1 (0.2) & 4.3 (0.1) \\
Schechter + Gaussian\tablenotemark{a} & -1.33 (0.04) & 3.1 (0.6)\tablenotemark{b}& 9.95 (0.05) & & \\
\\
2DSWML&-1.33 (0.02)& 4.8 (0.3) & 9.96 (0.02) & 4.3 (0.3) & 4.4 (0.1) \\
2DSWML, Non-Virgo&-1.34 (0.02)& 4.7 (0.3) & 9.96 (0.01) & 4.3 (0.3) & 4.4 (0.1) \\
\\
1/V$_{max}$, $\alpha.40_{1.8}$& -1.30 (0.03)& 4.6 (0.3)& 9.96 (0.02)& 4.0 (0.3)& 4.0 (0.1)\\
1DSWML, $\alpha.40_{1.8}$& -1.36 (0.06)& 4.5 (0.9)& 9.96 (0.04)& 4.4 (0.9)& 4.3 (0.3)\\
\\
HIPASS \citep{2005MNRAS.359L..30Z}\tablenotemark{c}& -1.37 (0.06)& 5 (1)& 9.86 (0.04) & 3.7 (0.5)&  \\
\\
Leo Group \citep{2009AJ....138..338S}\tablenotemark{d} & -1.41 (0.2) & & & &  \\
\hline
\enddata
\tablenotetext{a}{In the 1/V$_{max}$ case, pure Schechter functions provide a poor fit to the faint-end slope $\alpha$, which explains the difference in $\alpha$ for two fitting functions. The Gaussian component parameters are not shown in the table, given that they are not expected to be physical.}
\tablenotetext{b}{We defer to the 2DSWML measurement of $\phi_*$, due to the spurious feature in the 1/V$_{max}$ results.}
\tablenotetext{c}{Reported statistical and systematic errors combined in quadrature.}
\tablenotetext{d}{The excluded parameters $\phi_*$ and M$_*$ in the Leo Group are highly uncertain due to the lack of high-mass galaxies in its small volume.}
\end{deluxetable*}


\section{Discussion \label{discussion}}

\begin{figure}[ht!]
\epsscale{.8}
\plotone{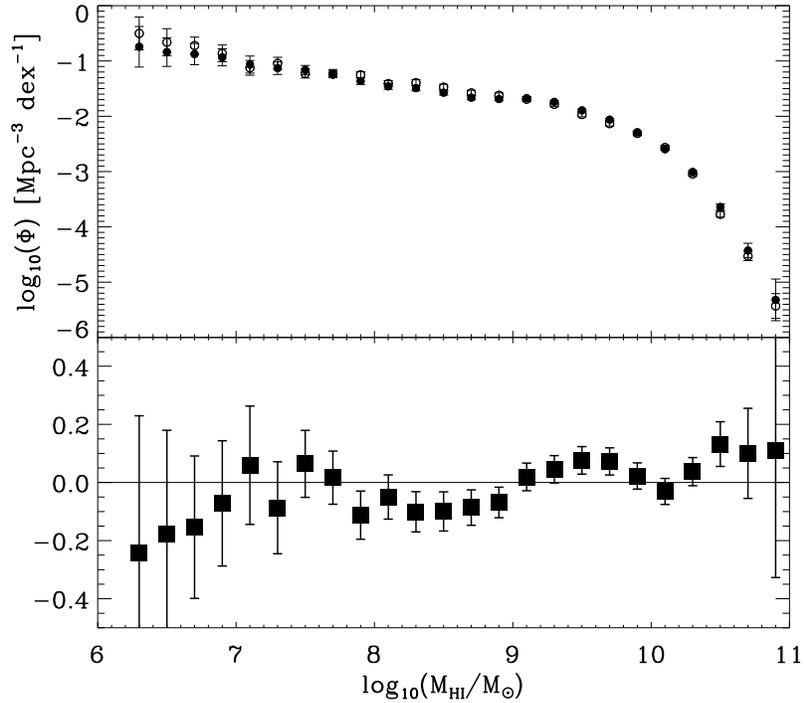}
\caption {Top panel: The HIMF derived from $\alpha.40$ with the 1/V$_{max}$ method (filled circles) and the 2DSWML method (open circles), with error bars. Bottom panel: The difference between the HIMF points, shown above, derived from the 1/V$_{max}$ and 2DSWML methods.\label{SWMLcompare}}
\end{figure}

Fig. \ref{SWMLcompare} compares the $\alpha.40$ HIMF derived via the $1/V_{max}$ method (filled circles) and the SWML method (open circles), and shows the difference between them in the bottom panel. The bin-by-bin differences between the SWML and 1/V$_{max}$ methods are small, and do not affect the measurement of $\Omega_{HI}$, though the faintest, most error-prone bins are found to be more populated in the SWML analysis. After we have corrected for the feature introduced to the 1/V$_{max}$ result by large-scale structure, we find excellent agreement between all measurements of $\alpha$ (-1.33 $\pm$ 0.02) and $\Omega_{HI}$ (4.3 $\pm$ 0.2).

In the case of 1/V$_{max}$, large-scale structure and the correction we estimate to deal with it have the largest impact on the final result.The 2DSWML method is designed to be insensitive to density fluctuations, and the agreement between the two measurements indicates that the large-scale structure correction is successful.

\subsection{Impact of the Virgo Cluster \label{excludevirgo}}

Measurements of the HI mass function can be sensitive to large-scale structure in the survey volume. As discussed above, we correct for large scale structure in the 1/$V_{max}$ method to ameliorate this effect, but our 2DSWML measurement could also be sensitive to this large nearby overdensity. To test the robustness of the 1/V$_{max}$ correction and of our derived HIMF, we consider the result obtained when we exclude the portion of $\alpha.40$ that crosses the Virgo cluster. Many of our low-mass objects are contributed by this nearby overdensity, and our large scale structure correction mechanism is the largest in this region; if we are correcting appropriately, we should obtain the same result regardless of the inclusion of the Virgo sources. This test is imperfect, given that the local volume generally is overdense. We exclude all galaxies lying within our adopted Virgo field, covering 12$^{h}$ $< \alpha <$ 13$^{h}$ and the full declination extent of the $\alpha.40$ survey \citep{2002MNRAS.333..423T}, reducing the sample size to $\sim$9200 for 1/V$_{max}$ and $\sim$ 8600 for 2DSWML. Errors are measured as described above, but in this case we jackknife resample over only 18 subregions.

Our results, within the errors, are the same whether or not we exclude the Virgo overdensity. This is true both for parameters and for our measurement of $\Omega_{HI}$. In the case of 1/V$_{max}$, we again find that a Schechter summed with a Gaussian provides a better fit to the data by accounting for features introduced by large-scale structure in the foreground of the Pisces-Perseus supercluster In Table \ref{params}, we compare our findings for samples inclusive and exclusive of Virgo. Additionally, we list the HIPASS HI mass function and the \citet{ 2009AJ....138..338S} HIMF of ALFALFA sources in the Leo group. In the case of the $\alpha.40$ and $\alpha.40_{1.8}$ samples, we also list the value of $\Omega_{HI}$ found by integrating the Schechter function fit or using the HIMF bin points. Each table entry is accompanied by 1$\sigma$ errors in parentheses.

\subsection{Comparison with Previous Work\label{prevresult}}

We find a value of $\Omega_{HI}$ that is 16$\%$ higher than the complete HIPASS survey value \citep{2005MNRAS.359L..30Z}. That HIPASS result is excluded by our $2\sigma$ errors, but the more preliminary HIPASS result \citep{2003AJ....125.2842Z} is in agreement with our result while carrying significantly larger error than we find. We also find 
$\log $ (M$_{*}$/M$_{\odot}$) = 9.96, so that the break in our HIMF occurs at masses 0.1 dex higher than was found in either of the HIPASS analyses. Since the high-mass end of the HIMF is sensitive to M$_{*}$, HIPASS significantly undercounts the highest-mass gas-rich galaxies. When our Schechter function is extrapolated to $\log $(M$_{*}/$M$_{\odot})$ = 11.0, we predict an order of magnitude more galaxies than HIPASS. At more modest values, $\log $ (M$_{*}$/M$_{\odot}$) = 10.75, this is reduced to a factor of $\sim$ 5.

\begin{figure}[ht!]
\epsscale{.7}
\plotone{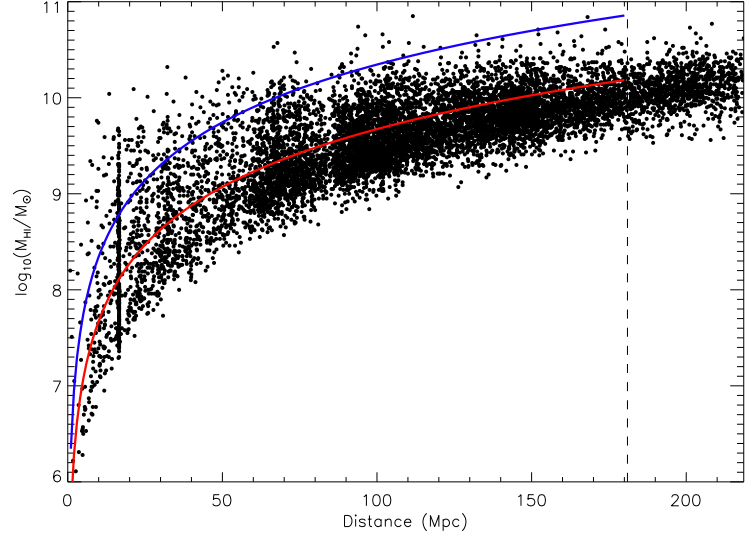}
\caption {$\alpha.40$ detections plotted as $log (M_{HI}/M_{\odot})$ vs. distance in Mpc. The upper (blue) solid line is the HIPASS completeness limit, and the lower (red) solid line is the HIPASS detection limit. The dashed vertical line shows the redshift limit of HIPASS assuming the ALFALFA adopted value H$_{0}$ = 70 \kms \ Mpc$^{-1}$.\label{spanhauer}}
\end{figure}

In Fig. \ref{spanhauer}, we show the mass of $\alpha.40$ detections as a function of their distance in Mpc, and compare that to the HIPASS completeness and detection limits. The dashed vertical line shows the 12,700 \kms \ redshift cutoff of HIPASS assuming H$_{0}$ = 70 km $s^{-1}$ Mpc$^{-1})$, demonstrating the ALFALFA survey's ability to probe the rare highest-mass galaxies at large redshifts. While the $\alpha.40$ sample extends only to 15,000 \kms / in order to avoid rfi, the full ALFALFA bandwidth allows us to probe to 18,000 \kms. Given that the survey was designed to be sensitive at those greater redshifts, we are still able to observe many galaxies at the limit of $\alpha.40$, while the \citet{2005MNRAS.359L..30Z} sample becomes very sparse near the survey's redshift limits.

This improved measurement of the HIMF has implications for work that relied upon the HIPASS results. Present-day HI surveys are limited in their ability to probe redshift space, even when they are targeted (z $<$ 0.5), so models of evolution of the HI mass function rely on the measurement at z = 0. Higher-precision measurements provide better constraints for evolutionary models. Numerical models of galaxy formation and evolution \citep{2010MNRAS.tmp..974P} depend on the z = 0 HIMF to assess the success of the models and to extrapolate that result to predictions for future HI surveys. For example, \citet{2010MNRAS.401..743A} predicted the ability of future HI line surveys with an instrument like the Square Kilometer Array (SKA) to constrain dark energy through measurements of the baryon acoustic oscillation scale. Those authors consider models of the HIMF evolution that are sensitive to the value M$^{*}$. Typically, these galaxy models also depend on the assumed H$_{2}$/HI ratio to convert simulated cold gas into atomic and molecular components (e.g. \citet{2004NewAR..48.1239B}), so updated estimates of either $\Omega_{HI}$ or $\Omega_{H_{2}}$ affect our ability to produce realistic models of gas-rich galaxies.

We confirm previous findings that $\Omega_{HI}$ at z = 0 is inconsistent with the value inferred from damped Lyman absorber (DLA) systems at z $\sim$ 2 and that significant evolution is required to reconcile measurements in the two epochs \citep{2009A&A...505.1087N,2006ApJ...636..610R}, while providing a tighter constraint on the present-day energy density of cold gas.

\subsection{Comparison with Simulations\label{simulations}}

\citet{2009ApJ...698.1467O} (hereafter O09) used the Millennium Simulation catalog, the \citet{2007MNRAS.375....2D} virtual catalog of galaxies, and a physically-motivated prescription to assign realistic gas (HI, He and H$_{2}$) masses at a range of redshifts. While this catalog has a limited ability to realistically trace detailed galaxy evolution and limited mass resolution -- down to about $10^{8.0} M_{\odot}$ of neutral hydrogen, which is comparable to the particle size in the Millennium run \citep{2005Natur.435..629S} -- it serves as the best currently available comparison of observed gas-rich disks with the underlying theory of dark matter halos. 

\subsubsection{Simulated HI Mass Function\label{simhimf}}

\begin{figure}[ht!]
\epsscale{0.7}
\plotone{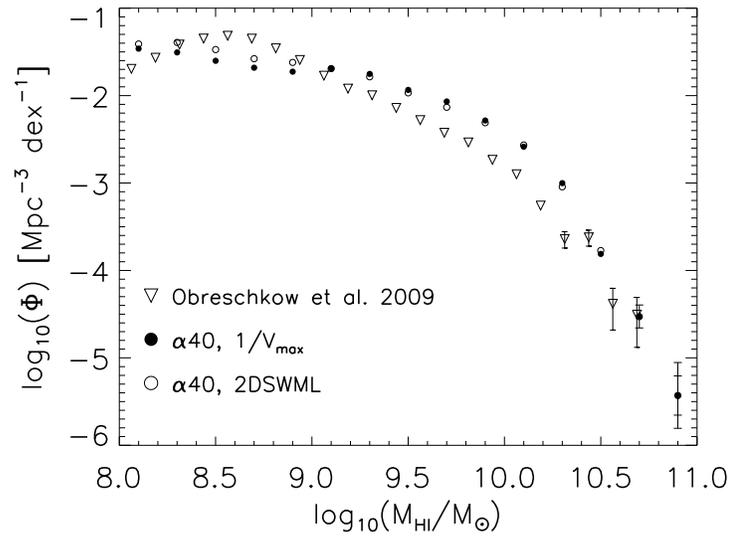}
\caption {The HIMF of the \citet{2009ApJ...698.1467O} analysis of cool gas in simulated galaxies from the Millennium run (open triangles), compared to the $\alpha.40$ 1/V$_{max}$ (filled circles) and 2DSWML (open circles) HIMFs. The ALFALFA sample is divided to 5 mass bins per dex, and the simulated galaxies to 8 bins per dex. Only the mass range $\log (M_{HI}/M_{\odot}) > 8.0$ is displayed, due to poor mass resolution in O09, and the simulated galaxy sample includes only galaxies at redshift z=0. For the ALFALFA HIMF, error bars represent both counting and mass estimate errors, but errors on the O09 HIMF are based on Poisson counting only. Where not visible, error bars are smaller than the plotted symbol size.\label{obreschkowHIMF}}
\end{figure}

O09 derive an HI mass function that is, in its gross properties, consistent with HIPASS \citep{2005MNRAS.359L..30Z}, ignoring spurious features near the mass resolution limit of the simulation. The O09 gas masses are obtained by combining the cold particle masses from the Millennium Run with a model to split the cold gas into molecular hydrogen and atomic hydrogen and helium components. Fig. \ref{obreschkowHIMF} compares the O09 HIMF, including only galaxies with $\log (M_{HI}/M_{\odot}) > 8.0$ and at redshift z=0, with the 1/V$_{max}$ and 2DSWML HIMFs derived in this work. $\Omega^{sim}_{HI}$ = 3.4 $\times 10^{-4}$ inferred from the O09 HIMF is in good agreement with this work and with HIPASS. While it is clear that the overall statistical distribution of the cold gas prescription generally recovers the overall density and the gross properties of the statistical distribution, the details of the O09 HIMF disagree with observations, particularly at the extreme low-mass end where the Millennium Run work suffers from poor resolution and inadequate merger histories. 

It is also worth noting that O09 report that they overpredict the number of high-mass sources in comparison to HIPASS, and suggest that this may be due to opacity in observed disks at these masses. However, we find that they underpredict high mass galaxies at z=0, the opposite effect. This is likely due to the O09 analysis of the HIMF, which is not limited to the final galaxies evolved to z=0; rather, their HIMF also includes galaxies at higher-redshift simulation snapshots which are presumably more gas-rich than their present-day counterparts. This would therefore overpredict the abundance of high-mass galaxies.

\subsubsection{Faint-End Slope\label{hoeft}}

As has been found in previous work, the faint-end slope of the $\alpha.40$ HIMF is significantly shallower than the Press-Schechter prediction of $\alpha \sim -1.8$ \citep{1974ApJ...187..425P}. Potentially, this difference can be linked to baryon loss and the suppression of accretion via photoheating in the low-mass dark matter halos. Simulations suggest that dark matter halos with masses below $\sim$ 6.5 $\times$ 10$^{9}$ h$^{-1}$ M$_{\odot}$ result in baryon-poor galaxies in present-day voids and other environments \citep{2008IAUS..244..279H,2006MNRAS.371..401H,2010arXiv1001.4721H}. In principle, the discrepancy could be explained by an argument invoking the mass scale at which photoheating becomes important.

A fitting function has been proposed \citep{2000ApJ...542..535G} to describe the behavior of baryon fraction as a function of underlying halo mass:

\begin{equation}
f_{b} = f_{b0}  \left [ 1 + (2^{\gamma/3} - 1) \left ( \frac{M_{c}}{M_{tot}} \right )^{\gamma} \right ]^{-3/\gamma}
\label{eqnfrac}
\end{equation}

\noindent where the parameters f$_{b0}$ and M$_{c}$ are, respectively, the baryon fraction in large halos and the characteristic halo mass where f$_{b}$=f$_{b0}$/2.

If decreasing baryon fraction with decreasing halo mass explains the difference between low-mass slopes in baryonic (stellar and HI) and halo mass functions, then this fitting function should consistently predict baryonic and cold gas mass functions with values of $\alpha \sim -1.3$. In the low-mass limit, the first term of Eqn. \ref{eqnfrac} can be dropped and the total mass in a halo can be assumed to be dominated by the dark matter, M$_{tot} \approx$ M$_{D}$. Via the definition f$_{B}$=M$_{B}$/M$_{D}$ we have

\begin{equation}
M_{D}=\frac{M_{B}}{f_{B}}=\frac{M_{B}}{f_{b0}} (2^{\gamma/3} - 1)^{3/\gamma}\left ( \frac{M_{D}}{M_{c}} \right )^{-3}
\label{eqnmd}
\end{equation}

Compressing all constants gives the relation $M_{D}\propto M_{B}^{1/4}$, which can then be used to relate the low-mass ends of the baryonic and dark matter mass functions. On the faint end of the dark matter mass function, the exponential term of the Schechter function can be dropped. From $\frac{d \log M_{D}}{d \log M_{B}}=\frac{1}{4}$ we can, finally, conclude that

\begin{equation}
\phi(M_{B}) = \frac{dn}{d \log M_{B}} \propto \ \phi_{*} \ \left (\frac{M_{B}}{M_{*,B}} \right )^{(\alpha_{D} + 1)/4} 
\label{eqnmbaryon}
\end{equation}

\noindent where $(\alpha_{D} + 1)/4=\alpha_{B} + 1$. Starting from the Press-Schechter prediction of a faint-end slope $\alpha_{D} \sim -1.8$, the consideration of baryon fraction leads to $\alpha_{B} \sim -1.2$, which is more consistent with HI and stellar mass functions. In principle, the discrepancy between dark matter simulations and observed baryon mass functions could be explained by the photoheating simulations of \citet{2006MNRAS.371..401H} and \citet{2010arXiv1001.4721H}.

The baryon fraction of O09's simulated galaxies loosely follows this descriptive baryon fraction function (Eqn. \ref{eqnfrac}). However, the halo mass scale at which the baryon loss starts to drop steeply is about two orders of magnitude larger than the scale found by the detailed hydrodynamical simulations of \citet{2008IAUS..244..279H}. Additionally, there is large scatter in the mass interval of interest, since the simulation's resolution is poor for the halo masses where baryon loss becomes important. The level of agreement between O09 and \citet{2008IAUS..244..279H} is therefore difficult to quantify, and we use the latter's determination of $f_{b}$ in what follows.

Eqn. \ref{eqnfrac} suggests that the baryonic content of low-mass galaxies in $\alpha.40$ may be severely biased with respect to the underlying halo mass distribution. If simulations accurately predict the relationship between initial halo masses and resulting baryon fractions after reionization and photoheating, then the application of $f_{b}$ should provide an estimate of the resulting baryon mass function at z=0. This depends on the extremely naive assumption that the cold HI gas content is depleted in the same fraction as the baryons overall.

The publicly available GENMF code\footnote{http://icc.dur.ac.uk/Research/PublicDownloads/genmf\_readme.html} produces halo mass function fits to the \citet{2007MNRAS.374....2R} N-body simulations at high resolution, from 10$^{5}$ to 10$^{12}$ h$^{-1}$ M$_{\odot}$. We adopt their mass function at z=0, with their suggested parameters $\Omega_M \approx$ 0.238, $\Omega_{\Lambda} \approx$ 0.762, and $\sigma_{8}$ = 0.74 (at z=0), and apply Eqn. \ref{eqnfrac} to extract the predicted baryon mass function and fit the faint-end slope. The results are displayed in Table \ref{baryon} for an exemplary set of values for $f_{b,0}$, $M_c$ and $\gamma$. 

Through this approach, it is possible to modify the underlying halo mass function ($\alpha \approx$ -1.8) to meet our observations ($\alpha \approx$ -1.3). The suggestion that low-mass halos may re-accrete cold gas at late times \citep{ 2009MNRAS.392L..45R}, if substantiated, could further change the shape of the resulting baryon mass function. While this approach indicates we may be close to resolving the missing satellites problem and the discrepancy between predicted and observed faint-end mass function slopes, the precise requirements of baryon depletion mechanisms are not well-constrained by available simulations.

\begin{deluxetable}{cccc}
\tablewidth{0pt}
\tabletypesize{\small}
\tablecaption{Faint-End Slopes of Modeled Baryon Mass Functions\label{baryon}}
\tablehead{ \colhead{$f_{b,0}$} & \colhead{$M_c$} & \colhead{$\gamma$} & \colhead{$\alpha$}\\
}
\startdata
0.20 & 9.0 & 1.0 & -1.30 \\
0.20 & 9.5 & 1.0 & -1.27 \\
0.16 & 9.0 & 1.0 & -1.31 \\
0.16 & 9.5 & 1.0 & -1.28 \\
0.16 & 9.0 & 1.5 & -1.24 \\
0.16 & 9.5 & 1.5 & -1.22 \\
0.16 & 9.0 & 2.0 & -1.21 \\
0.16 & 9.5 & 2.0 & -1.19 \\
0.15 & 9.0 & 1.0 & -1.31 \\
0.15 & 9.5 & 1.0 & -1.28 \\
0.15 & 9.0 & 2.0 & -1.21 \\
0.15 & 9.5 & 2.0 & -1.19 \\
\hline
\enddata
\end{deluxetable}

\section{Conclusions\label{conclusion}}

We have derived the HI mass function from a sample of $\sim$10,000 extragalactic sources comprising the ALFALFA 40$\%$ Survey, and have adapted the 1/$V_{max}$ method to fully account for survey sensitivity, large-scale structure, and mass errors. We have demonstrated the robustness of this method by testing flux-limited samples and by calculating the HIMF via a second approach, the structure-insensitive 2DSWML method. Our major result, the derivation of the global HIMF, indicates a Schechter function with parameters $\phi_{*}$ (h$_{70}^3$ Mpc$^{-3}$dex$^{-1}$) = 4.8 $\pm$ .3 $\times$ 10$^{-3}$, $\log $(M$_{*}/$M$_{\odot})$ + 2 $\log$ h$_{70}$  = 9.96 $\pm$ 0.02 and $\alpha$ = -1.33 $\pm$ 0.02. 

We find $\Omega_{HI}$= 4.3 $\pm$ 0.3 $\times 10^{-4}$ h$^{-1}_{70}$, a robust constraint that is 16$\%$ higher than the complete HIPASS survey value 3.7 $\times 10^{-4}$ h$^{-1}_{70}$ \citep{2005MNRAS.359L..30Z}, which we exclude at the $2\sigma$ level. The more preliminary HIPASS result \citep{2003AJ....125.2842Z} is in agreement with our result, but carries a significantly larger error. When we exclude the Virgo cluster from our analysis, the $\Omega_{HI}$ value remains stable, indicating that our measurements are robust against large-scale structure. In each case, we find the same value $\Omega_{HI}$ whether derived from the binned HIMF points themselves or from the best-fit Schechter parameters.

The larger values of $\Omega_{HI}$ and of M$_{*}$ that we find in comparison to HIPASS demonstrate ALFALFA's advantage in detecting high-mass galaxies at large distances. On the extreme high-mass end of the HI mass function, our measurement and the accompanying Schechter function predict an order of magnitude more galaxies at $\log $(M$_{HI}/$M$_{\odot}) \sim$ 11.0, and we find a factor of $\sim$5 more galaxies at $\log $(M$_{HI}/$M$_{\odot})$ = 10.75. This has implications for previous estimates of the detection rate of future large-scale HI line surveys with the SKA.

We confirm previous findings that significant evolution in cold gas reservoirs must occur between z $\sim$ 2 and z = 0 given that $\Omega_{HI}$ is a factor of $\sim$ 2 smaller in the former epoch compared with the latter \citep{2009A&A...505.1087N,2006ApJ...636..610R}. Further, we suggest that work on photoheating and other processes that prevent low-mass dark matter halos from accreting gas may be coming close to explaining the so-called `missing satellite problem' at low redshift. Further numerical work, particularly at resolutions capable of recovering low densities of cold gas at z=0, is required in this area of research.

Future work will consider the variation of the HI mass function with environment, and will include larger numbers of galaxies across a full range of extragalactic environments as the ALFALFA survey continues and new data products are released.

\acknowledgements
The authors would like to acknowledge the work of the entire ALFALFA collaboration team in observing, flagging, and extracting the catalog of galaxies used in this work.

This work was supported by NSF grants AST-0607007 and AST-9397661, and by grants from the National Defense Science and Engineering Graduate (NDSEG) fellowship and from the Brinson Foundation.

\bibliographystyle{apj}
\bibliography{myreferences}

\appendix

\section{Details of Corrections to the 1/V$_{max}$ Method\label{appA}}

\subsection{Width-Dependent Sensitivity Correction\label{widthcorrsec}}

\citet{2005AJ....130.2613G} predicted, from the precursor survey observations, that ALFALFA in full two-drift mode could expect an approximate integrated flux detection threshold, S$_{int, th}$ in Jy \kms, dependent upon profile width as follows:

\begin{equation}
   S_{int, th} = \left\{
     \begin{array}{lr}
       0.15 \, S/N \, (W_{50}/200)^{1/2}, \, W_{50} < 200 \\
       0.15 \, S/N \, (W_{50}/200), \, \, \, W_{50} \geq 200
     \end{array}
   \right.
\label{eqnthresh}
\end{equation}

In practice, however, ALFALFA outperforms this detection threshold, and we therefore use the data itself to fit a detection limit as described in \S \ref{corrs}. 

The width-dependent sensitivity correction is based on the distribution of observed profile widths. We also assume that the distribution of observed galaxies gives an indication of the true underlying distribution. We are therefore interested in working with as many sample galaxies as possible, and thus we consider a detection threshold S$_{int, th}$ as a function of W$_{50}$ that indicates the limits of ALFALFA's detection ability, rather than a strict completeness limit as in the 2DSWML case (\S \ref{swml}).

The completeness correction is based on the relationship of galaxy mass to the distribution of profile widths W$_{50}$. It is known that HI profile widths and masses are correlated, and we observe a mass-dependent spread in the distribution of profile width. We determine the profile width distribution as a function of mass by binning $\alpha.40$ galaxies by $\log (M_{HI}/M_{\odot})$ and fitting to each histogram a Gumbel (or Extreme Value Type 1) distribution:

\begin{equation}
f(x) = \frac{1}{\beta} \, e^{\frac{x-\mu}{\beta}} \, e^{-e^{\frac{x-\mu}{\beta}}}
\label{eqngumbel}
\end{equation}

\noindent where $\mu$ parametrizes the center of the distribution and $\beta$ its breadth. The profile width distributions feature narrow central peaks and extended skewed tails, which the Gumbel distribution is designed specifically to model.

We find that the center of the profile width distribution increases linearly with $\log (M_{HI}/M_{\odot})$, and the breadth decreases linearly with $\log (M_{HI}/M_{\odot})$. We derive a relationship between $\log (M_{HI}/M_{\odot})$ and the parameters $\mu$ and $\beta$, in order to extrapolate to any mass and infer the underlying distribution of W$_{50}$ to which a given galaxy belongs, $P(W_{50}, M_{HI})$. The probability of detecting a galaxy in a given mass bin depends on the profile width distribution for that bin, as well as the limiting profile width W$_{50,lim}$ beyond which that galaxy would not be detectable by ALFALFA. We are seeking a correction factor $C$ that will account for the profile width-integrated flux bias and that satisfies the relationship

\begin{equation}
N_{galaxies} (M_{HI}) = C \, N_{obs}(M_{HI})
\label{eqnprobs}
\end{equation}

\noindent where $N_{galaxies}$ is the corrected galaxy count to be input for the calculation of the HIMF, and $N_{obs}$ is the observed galaxy count. In terms of the derived distribution $P(W_{50}, M_{HI})$, we have

\begin{equation}
C = \frac{\int_{-\infty}^{+\inf} P (W_{50}, M_{HI}) \, dW_{50}}{\int_{-\infty}^{W_{50,lim}} P (W_{50}, M_{HI}) \, dW_{50}}
\label{eqnc}
\end{equation}

Since a bin is made up of galaxies with varying W$_{50,lim}$, we apply this correction to each individual galaxy, rather than on a mass bin-by-bin basis. The sum over effective search volume, $\Sigma 1/V_{max}$, therefore becomes $\Sigma C/V_{max}$.

To be conservative, we have included the errors on our derived linear relationships between $\log (M_{HI}/M_{\odot})$ and the Gumbel distribution parameters $\mu$ and $\beta$ in our final error analysis for the HI mass function.

\subsection{Large Scale Structure Correction\label{lss}}

The 1/$V_{max}$ method would be biased by large scale structure if we counted galaxies in overdense regions with the same weight as their counterparts in voids. Instead, we want to consider the effective search volume $V_{max, eff}$ in such a way that overdense regions are counted as contributing more effective volume to the overall survey.

We modify $\Sigma 1/V_{max}$ to include weighting by the average density $n(V_{max})$ interior to $D_{max}$, normalized to the average density of the Universe. The expression for measuring the HIMF then becomes $\Sigma 1/n(V_{max})V_{max}$ \citep{2005ApJ...621..215S}. We obtain $n(V_{max})$ from the PSCz density reconstruction of \citet{1999MNRAS.308....1B}, using their Cartesian map of evenly-spaced grid points out to 240 Mpc h$^{-1}$ smoothed to 3.2 Mpc h$^{-1}$ and using our assumed value h = 0.7. For values $D_{max} > \sim 85$ Mpc, the average density interior to $D_{max}$ becomes equal to the average density in the PSCz map, so no correction is needed. The large scale structure correction is therefore small compared to the Poisson counting error for galaxies with $\log (M_{HI}/M_{\odot}) > 9.0$, which are found at large distances.

This weighting scheme for galaxy counts in over- and under-abundant regions corrects the relative counts between different environments, so that clusters and superclusters don't dominate the shape of the measured HIMF.

\section{Details of the 2DSWML Method \label{appB}}

In the case of a sample such as $\alpha$.40, which is not flux-limited and instead depends on additional observables, we must consider a bivariate or two-dimensional stepwise maximum likelihood (2DSWML) approach. In this bivariate case, the likelihood of finding a galaxy with HI mass $M_{HI,i}$ and velocity width $W_{50,i}$ at distance $D_i$ is given by

\begin{eqnarray}
\ell_i = \frac{\phi(M_{HI,i},W_{50,i})}{\int_{W_{50}=0}^{\infty} \int_{M_{HI}=M_{HI,lim}(D_i,W_{50})}^{\infty} \: \phi(M_{HI},W_{50}) \: dM_{HI} dW_{50}}   \label{likelihood}
\end{eqnarray}

\noindent 
where $M_{HI,lim}(D_i,W_{50})$ is the minimum detectable mass at distance $D_i$ for a galaxy with velocity width $W_{50}$, calculated using the completeness relationship in integrated flux-velocity width space as described above. 

We proceed by splitting the distribution in bins of $m = \log(M_{HI}/M_\odot)$ and $w = \log W_{50}$, and assume a constant value within each bin. This leads to the Two-Dimensional Step Wise Maximum Likelihood (2DSWML) technique, where the parameters of the two-dimensional distribution can now be written as $\phi_{jk}$ ($j =1,2,...,N_m$ and $k=1,2,...,N_w$). The individual likelihood for each galaxy (Eqn. \ref{likelihood}) becomes 

\begin{eqnarray}
\ell_i = \frac{\sum_j \sum_k V_{ijk} \phi_{jk}}{\sum_j \sum_k H_{ijk} \phi_{jk} \Delta m \Delta w},  \label{swlikelihood}
\end{eqnarray}

\noindent
where the set of coefficients $V_{ijk}$ are used to ensure that only the value for the bin to which galaxy $i$ belongs appears in the numerator and the coefficients $H_{ijk}$ are used to enforce the summation in the denominator to go only over the area in the $(m,w)$ plane where galaxies could be detectable at distance $D_i$. More precisely,

\begin{eqnarray}
V_{ijk} = 
\left\{
\begin{array}{lr}
1& \mathrm{if}\: \mathrm{galaxy}\:i\: \mathrm{belongs}\: \mathrm{to}\: \mathrm{mass}\: \mathrm{bin}\: j \: \mathrm{and}\: \mathrm{width}\: \mathrm{bin}\: k\\
0& \mathrm{otherwise}
\end{array}
\right.
\end{eqnarray}  

\noindent
and, if we denote the completeness function in the $(m,w)$ plane for galaxies at distance $D_i$ by $C_i(m,w)$,   

\begin{eqnarray}
H_{ijk} = \frac{1}{\Delta m \Delta w} \int_{w_k^-}^{w_k^+} \int_{m_j^-}^{m_j^+} C_i(m,w) \: dm dw
\end{eqnarray}

\noindent
where $m_j^-$ and $m_j^+$ are the HI mass at the lower and upper boundary of mass bin $j$ correspondingly and similarly $w_k^-$ and $w_k^+$ are the upper and lower boundaries of width bin $k$. The completness function in the mass-width plane, $C_i(m,w)$, is directly derived from the $\alpha.40$ sample data, as in Fig. \ref{sensitivity}. For the 2DSWML method we restrict ourselves to galaxies above a strict completeness cut as a function of W$_{50}$, where the completeness is 1, excluding 321 galaxies ($\sim 3\%$ of $\alpha$.40) from the calculation of the mass function. 

The goal of the 2DSWML approach is to find the values of the parameters $\phi_{jk}$ that maximize the joint likelihood of finding all the galaxies in the sample simoultaneously, $\mathcal{L} = \prod_i \ell_i$. In practice it is more convenient to maximize the log-likelihood, which using Eqn. \ref{swlikelihood}, can be written as 

\begin{eqnarray}
\ln \mathcal{L} = \sum_i \ln \ell_i = \sum_i \sum_j \sum_k V_{ijk} \ln(\phi_{jk}\Delta m \Delta w) \nonumber \\ - \sum_i \ln \left(\sum_j \sum_k H_{ijk} \phi_{jk} \Delta m \Delta w\right) + \mathrm{const}.  
\end{eqnarray}

\noindent
$\ln \mathcal{L}$ is maximized by setting the partial derivatives with respect to each of the parameters equal to zero, giving

\begin{eqnarray}
\phi_{jk} = \frac{\sum_i V_{ijk}}{\sum_i \frac{H_{ijk}}{\sum_m \sum_n H_{imn} \phi_{mn}}} = \frac{n_{jk}}{\sum_i \frac{H_{ijk}}{\sum_m \sum_n H_{imn} \phi_{mn}}}   \label{solution}
\end{eqnarray}
 
\noindent
where $n_{jk}$ is the galaxy count in bin $j,k$. The Maximum Likelihood values for each parameter can be found by iterating Eqn. \ref{solution} until a stable solution is obtained. Finally, the HI mass distribution can be derived by the bivariate HI mass-velocity width distribution by marginalizing over velocity width, or

\begin{eqnarray}
\phi_j = \sum_k \: \phi_{jk} \: \Delta w .  \label{margin}
\end{eqnarray} 

Marginalizing the bivariate distribution over HI mass leads, instead, to the projected velocity width function for HI bearing galaxies, which will be the focus of a forthcoming publication.

As Eqns. \ref{likelihood} \& \ref{swlikelihood} imply, the overall normalization is lost in the process, and only the relative values of the parameters $\phi_{jk}$ are meaningful. Fixing the amplitude gives the HI mass function.

\subsection{HIMF Amplitude \label{amplitude}}
To transform the calculated probability density function into an HI mass function (e.g. transform the unitless $\{ \phi_k \Delta m \}$ into space densities) we evaluate the amplitude of the HIMF by matching the integral of the distribution to the inferred average density of galaxies in the survey volume $\bar{n}$, as in \citet{2003AJ....125.2842Z}. \citet{1982ApJ...254..437D} discuss various estimators for $\bar{n}$ that strike different balances between stability against poor knowledge of the selection function of the survey and immunity to large-scale structure. Since we believe we have a good understanding of the selection function out to $cz = 15000$ \kms, we choose to adopt the estimator that is least prone to bias, denoted by $n_1$, defined as

\begin{equation}
n_1 = V_{survey}^{-1} \int \frac{n(D) \: dD}{S(D)}  
\end{equation}

\noindent
where $n(D) \: dD$ is the number of galaxies in a spherical shell of thickness $dD$ and radius $D$, and $V_{survey}$ is the total survey volume. The selection function $S(D)$ is the fraction of galaxies detectable at distance $D$ and is given by

\begin{equation}
S(D) = \frac{\int_{w_{min}}^{w_{max}} \int_{m_{lim}(w,D)}^{m_{max}} \: \phi(m,w) \: dm \ dw}{\int \int \: \phi(m,w) \: dm \ dw} .  
\end{equation} 

\noindent
In the case of the 2DSWML method we evaluate $n_1$ by the expression

\begin{equation}
n_1 = V_{survey}^{-1} \sum_i \frac{1}{\sum_j \sum_k H_{ijk} \phi_{jk} \Delta m \Delta w}.  \label{n1}
\end{equation}

\noindent
Eqn. \ref{n1} corresponds to weighing each detected galaxy in the survey by the inverse of the selection function at the galaxy's distance, effectively correcting each detection by the fraction of galaxies that cannot be detected at distance $D_i$.

\end{document}